\documentclass[12pt]{article}
\usepackage{a4wide,epsfig,psfrag,amsmath,amssymb,cite,scalefnt,wasysym}
\usepackage{color}
\usepackage{makeidx}
\usepackage{array}
\usepackage{longtable}
\usepackage{slashed}

\definecolor{Black}{RGB}{0,0,0}
\definecolor{SkyBlue}{RGB}{200,225,255}

\newcommand{\gsim}{\;\rlap{\lower 3.5 pt \hbox{$\mathchar \sim$}} \raise 1pt
 \hbox {$>$}\;}
\newcommand{\lsim}{\;\rlap{\lower 3.5 pt \hbox{$\mathchar \sim$}} \raise 1pt
 \hbox {$<$}\;}

\allowdisplaybreaks

\makeindex

\begin{document}

\title{\vskip-3cm{\baselineskip14pt
    \begin{flushleft}
      \normalsize HU-EP-19/35\\
  \end{flushleft}}
  \vskip1.5cm
  Fermion Traces Without Evanescence\\ 
}

\author{
  Nikolai Zerf
  \\[1em]
  {\small\it Institut f\"ur Physik}\\
  {\small\it Humboldt-Universit\"at zu Berlin}\\
  {\small\it Netwonstra{\ss}e 15}
  {\small\it D-12489 Berlin, Germany}
}

\date{}

\maketitle

\thispagestyle{empty}

\begin{abstract}
We outline the evaluation of $n$-dimensional fermion traces ($n\in\mathbb{N}$) built by products of Dirac-$\gamma$ matrices suitable for a uniform dimensional continuation.
Such a continuation is needed for calculations employing a dimensional regulator whenever intrinsically integer dimensional tensors yield non-vanishing contributions.
A prime example for such a tensor is given by $\gamma_5$ for $n=4$.
The main difference between Dimensional Regularization (DREG) and a Dimensionally Continued Regularization (DCREG)
is that DCREG does not attempt to lift the algebra to continuous $d$ dimensions ($d\in\mathbb{R}$).
As a consequence one has to properly deal with evanescent structures in order to ensure the uniform application of the regulator.
In basic steps we identify evanescent structures in fermion traces
and show that their proper treatment is crucial for example when calculating the $VVA$ anomaly in four dimensions.
We checked that the performed considerations enable the evaluation of Standard Model (SM) $Z$-factors within DCREG up to including three loops.

 \medskip

  \noindent

\end{abstract}

\thispagestyle{empty}


\newpage

\section{Introduction}\label{SEC:INTRO}
\enlargethispage{0.5cm}
Dimensional Regularization (DREG)~\cite{tHooft:1972tcz} is a very powerful regularization scheme that keeps internal and external symmetries like Lorentz-symmetry intact, 
without increasing the number of scales present in Feynman integrals,
as is the case for example in a Pauli-Villars regularization.
This keeps the appearing integrals simple enough to perform amplitude evaluation at the multi-loop order.

A commonly accepted application of DREG is however, restricted to the case where the amplitudes considered do not contain contributions made up by intrinsically integer dimensional tensors like $\varepsilon$-tensors,
because in this case one can safely lift the algebra to a general non-integer dimension $d$.
Basically this continuation to $d$ dimensions is then already performed at the level of Lagrangians.
Doing so one greatly benefits from the fact that existing Ward-identities are promoted to $d$ dimensions and ensures that bare results do not violate them.
It further ensures that the continuation to $d$ dimensions is automatically carried out in a uniform way.
There are many example applications of such a lifting to $d$ dimensions across various integer dimensions $n$.
Further it has been shown for many cases, that the regulated amplitudes are very suitable for the process of renormalization and infrared subtractions.
For a recent review of dimensional regulators and schemes see Ref.~\cite{Gnendiger:2017pys}.

However, in case one encounters an intrinsically integer dimensional tensor structure in an amplitude, 
a generically applicable dimensional scheme is not well established. 
Meaning for specific cases there exist treatments of $\varepsilon$-tensors that are known to not work for the general case.
Specifically there are many known ways (see Ref.~\cite{Gnendiger:2017rfh}) how to treat $\varepsilon$-tensors at the one-loop level, because one encounters for example only global UV divergences.
In more detail, there is no need for a renormalization of sub-diagrams.

For calculations beyond the one-loop order there have been general proposals how to treat $\varepsilon$-tensors within DREG.
However, to the best of the author's knowledge, there is no prescription available that leads to unambiguous results for an arbitrary physical observable.
In fact the lack of a consistent treatment of $\varepsilon$-tensors within DREG is the main reason
why up to now the Standard Model (SM) $\beta$-functions are fully known at three-loop order~\cite{Mihaila:2012fm,Mihaila:2012pz,Chetyrkin:2012rz,Chetyrkin:2013wya,Bednyakov:2013eba}, 
but only partially known up to four loops~\cite{Bednyakov:2015ooa,Zoller:2015tha,Chetyrkin:2016ruf},
although the computational abilities nowadays allow for a five-loop evaluation for example of the QCD $\beta$-function~\cite{Baikov:2008cp,Baikov:2010je,Baikov:2012zm,Baikov:2014qja,Chetyrkin:2016uhw,Baikov:2016tgj,Luthe:2016ima,Luthe:2016xec,Herzog:2017ohr,Luthe:2017ttc,Baikov:2017ujl,Luthe:2017ttg,Chetyrkin:2017bjc}.

In this context one has to mention that it has already been proven that a dimensional regulator can be used for a dimensional renormalization in the presence of $\gamma_5$ including the aforementioned $\varepsilon$-tensor contributions for arbitrary loop orders in Ref.~\cite{Breitenlohner:1977hr}.
However, the employed treatment of fermion traces known as the 't Hooft-Veltman scheme requires us to distinguish intrinsically four and $d$ dimensional vectors and it is known to require Ward-identity restoring counter terms.
The first point is for an actual multi-loop calculation feasible but very costly, because it tremendously increases the number of terms.
The second point is an actual conceptual problem at the multi loop level, which is not discussed at all in the publication.
One has to figure out how to systematically obtain the required subtraction terms.
We claim that it is impossible to systematically subtract the required terms at the integrand level without breaking the used algebra, 
because it in general requires the inclusion of evanescent terms, as we will see in Sec.~\ref{SEC:TRDC}.

\enlargethispage{0.5cm}

An explicit application of DREG to amplitudes involving a single $\gamma_5$ was carried out for the computation of the 3-loop QCD corrections to the pseudo scalar\footnote{See the preprint version.} and axial-vector current~\cite{Larin:1993tq}.
However, it is well known that the method used involving Ward-identity restoring finite subtraction terms cannot be applied to a general problem involving $\gamma_5$,
because the amplitude is required to have a special structure in order to allow for integrated, finite overall subtraction terms.
All of which is not discussed at all in Ref.~\cite{Larin:1993tq}. However it provides a very clear description of the carried out procedure.

We do not intend to give a complete review of the existing literature dealing with the problem of $\gamma_{5}$ in DREG,
but just reference a selection~\cite{Chanowitz:1979zu,Kreimer:1989ke,Korner:1991sx,Dugan:1990df,Kreimer:1993bh,Trueman:1995ca,Jegerlehner:2000dz,Moch:2015usa,Anger:2018ove}.

However, none of the papers provides a self consistent all order/all process prescription for the treatment of intrinsically integer dimensional tensors
such that a dimensional regularization at multi-loop order becomes viable.

In this article we start from scratch and restrict ourselves to the evaluation of fermion traces, only.
Here we identify evanescent contributions as a potential threat to the uniformity of trace results when dealing with intrinsically integer dimensional tensors in the dimensional regularization framework.
We show that a consistent elimination of these contributions can remove this threat.

This article is organized as follows:
We introduce our notation and remind the reader of important properties of the vector and spin representation of $SO(N)$ in Sec.~\ref{SEC:BASICS}.
We recommend Ref.~\cite{Cvitanovic:2008zz} for a more complete introduction to group theory aspects.

In Sec.~\ref{SEC:DC} we introduce the concept of dimensional continuation of fermion trace results and the concept of evanescent tensors/contributions.

In Sec.~\ref{SEC:MAXDEC} we systematically find all evanescent contributions within trace results evaluated for arbitrary dimension employing a maximal decomposition.

In Sec.~\ref{SEC:TRDC} we elaborate how to deal with different kinds of vector index types -- continuous and integer dimensional ones -- 
in order to preserve the symmetries of the corresponding integer dimensional space during a trace reduction.

In Sec.~\ref{SEC:SINGLEGAMMA5ANDEVANS} we show that evanescent contributions are able to break the expected symmetries of integer dimensional traces and thus their elimination is required.

In Sec.~\ref{SEC:REDPRES} we give a blueprint for a fermion trace reduction that yields results that can be used for a dimensionally continued regularization.

In Sec.~\ref{SEC:VVAANOMALY} we recalculate the well known $VVA$ anomaly arising in four space-time dimensions using our prescription.

We use Sec.~\ref{SEC:DISCUSSION} to discuss the implications following from our analysis and conclude with a summary in the last section.

\section{Basic Properties of $SO(N_V)$ Vectors \& Spin }\label{SEC:BASICS}
\subsection{Basic Relations Involving the Vector Representation}
In the following we are mainly dealing with the algebra and elements of the special orthogonal Lie group $SO(N_V)$.
All elements of the groups keep a symmetric and real tensor of rank two invariant, the metric tensor $\eta$,  meaning:
\begin{align}
 G^T(\omega)\cdot \eta \cdot G(\omega)= G(\omega)\cdot \eta \cdot G^T(\omega) = \eta\,.
\end{align}
Here $G(\omega)$ is an element of the group.
For convenience we just require $\eta$ to be Euclidean\footnote{One can continue all our results to be applicable for Minkowski case.} and obey:
\begin{align}
  \eta \cdot \eta =\eta\,.
\end{align}
Further we introduce the vector representation indices $\{\mu,\nu,\dots\}$ which run from $1$ to $N_V$.
We can then write the above invariance condition of the metric tensor following Einstein's summation convention as:
\begin{align}
 G_{\mu}^{\phantom{\nu}\nu}(\omega) G_{\sigma}^{\phantom{\nu}\rho}(\omega) \eta_{\nu\rho}= \eta_{\mu\sigma} \,.
\end{align}
Note that upper and lower indices do not transform differently, but are only used to indicate a proper implicit summation.
The trace of the metric tensor is given by the dimension of the irreducible vector representation:
\begin{align}
 \eta_{\mu\nu}\eta^{\nu\mu}= N_V\,.
\end{align}
Investigating the transformation properties of the dyadic product $\eta\otimes\eta$ reveals
that it can be decomposed into three different irreducible representations (irreps).
The decomposition can be described by:
\begin{align}
 V\otimes V=1\oplus\hat{V}_2\oplus\tilde{V}_2\,.
\end{align}
On projector level the decomposition reads explicitly:
\begin{align}
\eta^{\nu_1}_{\mu_1}\eta^{\nu_2}_{\mu_2}=\frac{1}{N_V}\eta^{\nu_1 \nu_2}\eta_{\mu_1 \mu_2}+\tfrac{1}{2}\left(\eta^{\nu_1}_{\mu_1}\eta^{\nu_2}_{\mu_2}-\eta^{\nu_1}_{\mu_2}\eta^{\nu_2}_{\mu_1}\right)+\left[\tfrac{1}{2}\left(\eta^{\nu_1}_{\mu_1}\eta^{\nu_2}_{\mu_2}+\eta^{\nu_1}_{\mu_2}\eta^{\nu_2}_{\mu_1}\right)-\frac{1}{N_V}\eta^{\nu_1 \nu_2}\eta_{\mu_1 \mu_2}\right]\,.
\end{align}
Here the first term is the projector on the singlet irrep ($1$) and the second term is the projector on the anti-symmetric two vector index irrep ($\hat{V}_2$).
The third term projects onto the traceless symmetric two vector index irrep ($\tilde{V}_2$).
The dimension of each representation can be calculated by taking the trace of each projector by identifying the indices $\mu_1$ with $\nu_1$ and $\mu_2$ with $\nu_2$.
One obtains the dimensions:
\begin{align}
 N_V \times N_V = 1 + \tfrac{1}{2}N_V(N_V-1) + \tfrac{1}{2}(N_V+2)(N_V-1)\,.
\end{align}
It turns out that the the anti-symmetric two vector index irrep is in fact the adjoint representation with dimension $N_A=N_{\hat{V}_2}=\tfrac{1}{2}N_V(N_V-1)$.
To see this we first write the group elements in terms of exponentiated generators $G(\omega)={\rm exp} (i T_V \cdot \omega)$ and expand them for small transformations:
\begin{align}
 G(\omega)_{\mu}^{\phantom{\nu}\nu}\approx  \eta_{\mu}^{\nu} + i\left(T^a_V\right)_{\mu}^{\phantom{\nu}\nu}\omega^a\,.
\end{align}
In linear order of any $\omega$ we obtain the following invariance condition for $\eta$:
\begin{align}
\left(T^a_V\right)^T \cdot \eta + \eta \cdot T^a_V=0\,,
\end{align}
Multiplying from left and right with $\eta$ yields:
\begin{align}
\eta \cdot\left[\left(T^a_V\right)^T+ T^a_V\right]\cdot \eta=0\,.
\end{align}
This equation is fulfilled as long as the generators are anti-symmetric under the exchange of the two vector indices:
\begin{align}
 \left[\left(T^a_V\right)^T\right]_{\mu\nu} =\left(T^a_V\right)_{\nu\mu}= -\left(T^a_V\right)_{\mu\nu}\,.
\end{align}
Because the dimension of the anti-symmetric two vector index irrep is given by $N_A$, the adjoint index $a$ runs from $1$ to $N_A$.
We further have the reduction identity:
\begin{align}
\sum\limits_{a=1}^{N_A} \left(T^a_V\right)^{\nu_1 \nu_2}\left(T^a_V\right)_{\mu_1 \mu_2} = \frac{I_{2,V}}{2}\left[\eta^{\nu_1}_{\mu_1}\eta^{\nu_2}_{\mu_2}-\eta^{\nu_1}_{\mu_2}\eta^{\nu_2}_{\mu_1}\right]   \,.
\end{align}
Here $I_{2,V}$ is called the second index of the vector representation and it depends on the normalization of the generators $T^a_V$.
A convenient choice is $I_{2,V}=2$.
We further see that the generators perform the transition into an eigenbasis of the adjoint subspace.
Their entries are special cases of Clebsch-Gordan coefficients.

In the following we will be using the irrep of $n$ fully anti-symmetric vector indices $\hat{V}_n$.
The projector from the $n$ vector index space onto $\hat{V}_n$ is given by the anti-symmetrizer
$A_{n}(\nu_1,\dots\nu_n;\mu_1,\dots,\mu_n)$
which is a vector index tensor of rank $2n$. 
The anti-symmetrizer is fully anti-symmetric under the permutation of the first and last $n$ vector indices.
Further it is symmetric under the full exchange of the first $n$ with the last $n$ indices.
For example:
\begin{align}
 A_{n}(\nu_1,\nu_2,\dots\nu_n;\mu_1,\dots,\mu_n)=&-A_{n}(\nu_2,\nu_1,\dots\nu_n;\mu_1,\dots,\mu_n)\,,\\
 A_{n}(\nu_1,\dots\nu_n;\mu_1,\mu_2,\dots,\mu_n)=&-A_{n}(\nu_1,\dots\nu_n;\mu_2,\mu_1,\dots,\mu_n)\,,\\
  A_{n}(\nu_1,\dots\nu_n;\mu_1,\dots,\mu_n)&= A_{n}(\mu_1,\dots,\mu_n;\nu_1,\dots\nu_n)\,.
\end{align}
Further it fulfills the projector condition $A_n\cdot A_n=A_n$:
\begin{align}
 A_{n}(\nu_1,\dots\nu_n;\mu_1,\dots,\mu_n)A_{n}(\mu_1,\dots,\mu_n;\rho_1,\dots\rho_n)=A_{n}(\nu_1,\dots\nu_n;\rho_1,\dots\rho_n)\,.
\end{align}
As special cases we have:
\begin{align}
 A_0=&1\,,\\
 A_1(\nu;\mu)=&\eta^{\nu}_{\mu}\,,\\
 A_2(\nu_1,\nu_2;\mu_1,\mu_2)=&\tfrac{1}{2}\left(\eta^{\nu_1}_{\mu_1}\eta^{\nu_2}_{\mu_2}-\eta^{\nu_1}_{\mu_2}\eta^{\nu_2}_{\mu_1}\right)\,.
\end{align}
Higher rank $A_n$'s can be recursively defined through lower rank ones using:
\begin{align}
 A_{n}(\mu_1,\dots,\mu_{n};\nu_1,\dots,\nu_{n})=\frac{1}{n}\Big[&+\eta^{\mu_1}_{\nu_1} A_{n-1}(\mu_2,\dots,\mu_{n};\nu_2,\nu_3,\dots,\nu_{n})\nonumber\\
 &-\eta^{\mu_1}_{\nu_2} A_{n-1}(\mu_2,\dots,\mu_{n};\nu_1,\nu_3,\dots,\nu_{n})\nonumber\\
 &+\dots\nonumber\\
 &+(-1)^n\eta^{\mu_1}_{\nu_n} A_{n-1}(\mu_2,\dots,\mu_{n};\nu_1,\dots,\nu_{n-1})\Big]\,.\label{EQ:RECANDEF}
\end{align}
In total there are $n$ terms inside the bracket on the r.h.s.\ of the equation.
The sign of the terms in the sum is alternating when ordered as indicated.
From the recursive construction it immediately follows that $A_n$ contains $n!$ terms.

The dimension of the irrep $\hat{V}_n$ can be calculated via the trace of the corresponding projector operator:
\begin{align}
 N_{\hat{V}_n}=A_{n}(\mu_1,\dots\mu_n;\mu_1,\dots,\mu_n)=\left(\begin{array}{l}N_V\\ n\end{array}\right)=\frac{N_V!}{n!(N_V-n)!}\,.
\end{align}
In case of non-integer dimension $N_V$ this expression can be analytically continued using the gamma function via $m!\rightarrow \Gamma(m+1)$.
Beside the anti-symmetrizers $A_n$ we can also define the symmetrizers $S_n$ which are totally symmetric under permutation of representation indices.
However, the $S_n$ do not project onto irreducible representations of the $SO(N)$ groups because they have a non-vanishing overlap with the singlet
(see the decomposition of $V\otimes V$ above).

Besides keeping $\eta$ invariant the group elements further have the property that their determinant is given by ${\rm det }[G(\omega)]=+1$.
The determinant can be written by:
\begin{align}
 A_{N_V}(\nu_1,\dots,\nu_{N_V};\mu_1,\dots,\mu_{N_V}) G(\omega)_{\mu_1}^{\phantom{\nu}\nu_1} \dots G(\omega)_{\mu_{N_V}}^{\phantom{\nu}\nu_{N_V}}=+1.\label{EQ:DETgEQ1}
\end{align}
When $N_V$ takes integer values $N_V> 1$ this condition induces another invariant tensor of vector index rank $N_V$.
In order to see this, one has to notice that the dimension of the irrep $\hat{V}_{N_V}$ in this case is given by $N_{\hat{V}_{N_V}}=1$ and thus is a singlet.
We are then able to write the projector with the help of the following factorizing Clebsch-Gordan coefficients:
\begin{align}
 A_{N_V}(\nu_1,\dots,\nu_{N_V};\mu_1,\dots,\mu_{N_V})= \tfrac{1}{N_V!}[\varepsilon_{N_V}]_{\nu_1,\dots,\nu_{N_V}}~[\varepsilon_{N_V}]^{\mu_1,\dots,\mu_{N_V}}\,.\label{EQ:AnSplitTo2eps}
\end{align}
We choose the overall normalization of $1/N_V!$ so that the entries of $\varepsilon_{N_V}$ have values of $\pm 1$ and $0$, only.
In more detail we have:
\begin{align}
 [\varepsilon_{N_V}]^{1 2\dots N_V}=[\varepsilon_{N_V}]_{1 2 \dots N_V}=1\,,
\end{align}
Requiring the indicated transformation properties of the new totally anti-symmetric tensor of rank $N_V$ to hold:
\begin{align}
 G(\omega)_{\mu_1}^{\phantom{\nu}\nu_1} \dots G(\omega)_{\mu_{N_V}}^{\phantom{\nu}\nu_{N_V}}[\varepsilon_{N_V}]_{\nu_1\dots\nu_{N_V}} =[\varepsilon_{N_V}]_{\mu_1\dots\mu_{N_V}}\,,
\end{align}
reduces the l.h.s. of Eq.~(\ref{EQ:DETgEQ1}) exactly to one after using Eq.~(\ref{EQ:AnSplitTo2eps}), 
because all group elements are absorbed by one $\varepsilon_{N_V}$ leaving just $A_{N_V}(\mu_1,\dots,\mu_{N_V};\mu_1,\dots,\mu_{N_V})$ the dimension of $\hat{V}_{N_V}=+1$.

\subsection{The Clifford Algebra for Spin Representations}
Besides the vector representation discussed so far one encounters spinor representations of $so(N_V)$ when
investigating the roots of the Klein-Gordon equation:
\begin{align}
 (\partial_{\mu}\partial^{\mu} + m^2)\phi=0\,.
\end{align}
Dirac introduced the Dirac matrices $\gamma^\mu$ (with $\slashed \partial=\partial_{\mu} \gamma^{\mu}$ )
to obtain a differential equation of first order for the spinor wave function $\Psi$.
\begin{align}
 (i\slashed \partial-m)\Psi=0\,.
\end{align}
When the Dirac matrices obey the Clifford-algebra 
\begin{align}
\{\gamma^{\mu},\gamma^{\nu}\}=2\eta^{\mu\nu}\,,\label{EQ:DA}
\end{align}
then each spinor component $\Psi_i$ ($i\in\{1,\dots ,N_S\}$) obeys the Klein-Gordon equation:
\begin{align}
 \left[(i\slashed \partial+m)(i\slashed \partial-m)\Psi\right]_i=-(\partial_{\mu}\partial^{\mu} + m^2)\Psi_i=0\,.
\end{align}
And the dimension of the spinor space is given by $N_S$.

\begin{figure}
\begin{tabular}{c}
 \includegraphics{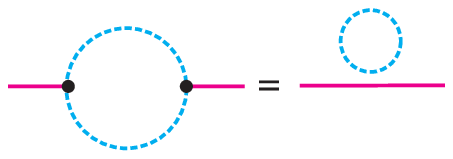}\\
  \includegraphics{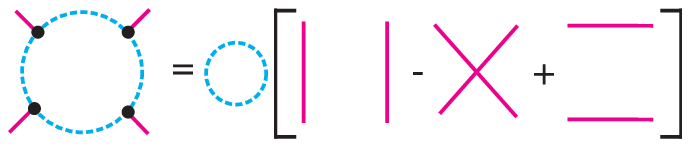}\\
  \includegraphics{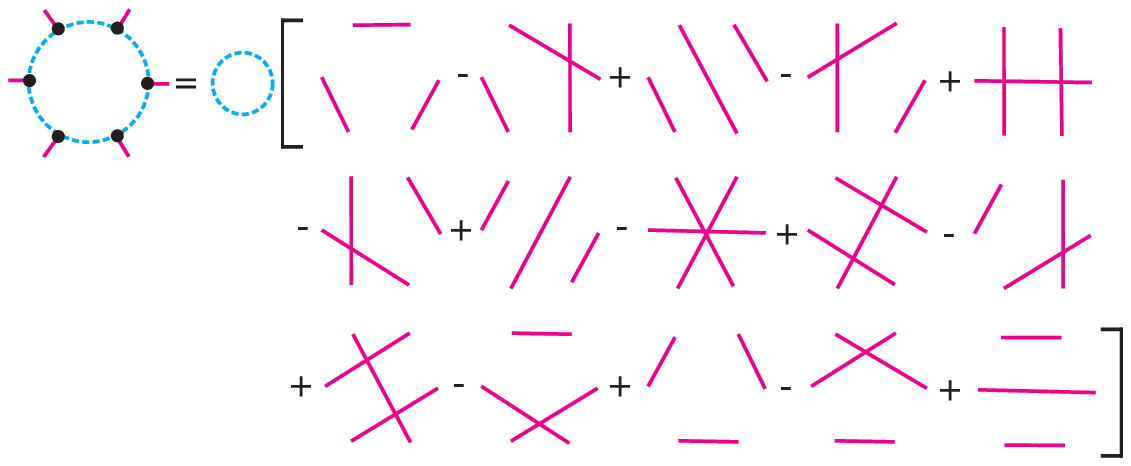}
\end{tabular}
\caption{\label{FIG:TrN}Results for $T_N$ in pictorial birdtracks notation. 
Dashed (cyan) lines represent spin contractions. 
Each black dot represents a Dirac-matrix.
A closed loop formed by a dashed (cyan) line represents the trace of the unit matrix in spinor space and is thus equal to $N_S$. 
Every solid (magenta) line denotes a metric $\eta$ tensor.
An introduction to the birdtracks notation is given in Ref.~\cite{Cvitanovic:2008zz}}
\end{figure}

The evaluation of traces made up by products of Dirac matrices is an immediate consequence of dealing with quantum states transforming under spinor representations of $SO(N_V)$.
We are thus interested in evaluating any trace of $N$ Dirac matrices:
\begin{align}
 T_{N}^{\mu_1\dots\mu_N}={\rm tr}\{\gamma^{\mu_1}\cdots\gamma^{\mu_N}\}\,.
\end{align}
In order to do so, we use Eq.~(\ref{EQ:DA}) to anti-commute the first $\gamma$ matrix through a chain of $N-1$ $\gamma$ matrices.
Each anti-commutation adds the value of the anti-commutator on the r.h.s:
\begin{align}
 \gamma^{\mu_1}\cdots\gamma^{\mu_N}=&2\eta^{\mu_1 \mu_2}\gamma^{\mu_3}\cdots\gamma^{\mu_N}+\dots+(-1)^{N} 2\eta^{\mu_1 \mu_N}\gamma^{\mu_2}\cdots\gamma^{\mu_{N-1}}\nonumber\\
  &+(-1)^{N+1}\gamma^{\mu_2}\cdots\gamma^{\mu_N}\gamma^{\mu_1}\,.\label{EQ:AntiCommutingThroughgammas}
\end{align}
There are $N-1$ terms with alternating signs in the first line.
Sorted in the indicated order the  metric tensor of the $n$-th term carries the index $\mu_1$ and $\mu_{n+1}$.
We immediately see that the obtained result yields an iterative trace reduction algorithm in case $N=2n$ is even, because we can use the cyclicity of the trace:
\begin{align}
{\rm tr}\{\gamma^{\mu_1}\cdots\gamma^{\mu_{2n}}\}=&\tfrac{1}{2}{\rm tr}\{\gamma^{\mu_1}\gamma^{\mu_2}\cdots\gamma^{\mu_{2n}} + \gamma^{\mu_2}\cdots\gamma^{\mu_{2n}}\gamma^{\mu_1}\}\nonumber\\
=&{\rm tr}\{\eta^{\mu_1 \mu_2}\gamma^{\mu_3}\cdots\gamma^{\mu_N}+\dots+\eta^{\mu_1 \mu_{2n}}\gamma^{\mu_2}\cdots\gamma^{\mu_{2n-1}}\}\,.
\end{align}
The obtained equation allows us to express traces of products of $N=2n$ Dirac matrices by traces of $2n-2$.
For the first four non-trivial cases one gets:
\begin{align}
  T_{2}^{\mu_1\mu_2}=& \eta^{\mu_1 \mu_2} {\rm tr}\{ 1_{N_S\times N_S}\}=N_S \eta^{\mu_1 \mu_2}\,,\\
  T_{4}^{\mu_1\mu_2\mu_3\mu_4}=& N_S\left(\eta^{\mu_1 \mu_2}\eta^{\mu_3 \mu_4} -\eta^{\mu_1 \mu_3}\eta^{\mu_2 \mu_4}+\eta^{\mu_1 \mu_4}\eta^{\mu_2 \mu_3}\right)\,,\\
  T_{6}^{\mu_1\mu_2\mu_3\mu_4\mu_5\mu_6}=&N_S\left(\eta^{\mu_1 \mu_2}\eta^{\mu_3 \mu_4}\eta^{\mu_5 \mu_6}+ \dots[14~\text{terms}]\right)\,,\label{EQ:T6}\\
  T_{8}^{\mu_1\mu_2\mu_3\mu_4\mu_5\mu_6\mu_7\mu_8}=&N_S\left(\eta^{\mu_1 \mu_2}\eta^{\mu_3 \mu_4}\eta^{\mu_5 \mu_6}\eta^{\mu_7 \mu_8}+ \dots [104~\text{terms}]\right)\,.
\end{align}
A pictorial representation can be found in terms of birdtracks in Fig.~\ref{FIG:TrN}.

It turns out that $T_{2n}$ reduces to $(2n-1)!!$ terms, where each term is a tensor product of $n$ metric tenors.
Each term has a unique pairing of the $2n$ vector indices.
There are in total  $(2n-1)!!$ ways of pairing $2n$ indices in products of metric tensors.
That means the trace contains all possible ways of pairing $2n$ indices in products of metric tensors.
As a consequence the trace does not depend on the reading direction within a trace, meaning:
\begin{align}
 T_{2n}^{\mu_1\mu_2\dots \mu_{2n}}=&T_{2n}^{ \mu_{2n}\dots\mu_2\mu_1}\,.\label{EQ:NOReadingDirecion}
\end{align}
The number of terms generated by a trace grows rapidly with $n$ like the following table shows.
\begin{center}
 \begin{tabular}{c|c}
 $n$& $(n-1)!!$\\
 \hline
 $2$ & $1$\\
  $4$ & $3$\\
  $6$ &$15$\\
  $8$ & $105$\\
  $10$ & $945$\\
  $12$ & $10395$
\end{tabular}
\end{center}
The sign of each term is given by the signature $\mathcal{S}$ of the permutation of the indices.
We have for example:
\begin{align}
 \mathcal{S}(\eta^{\mu_1 \mu_2}\eta^{\mu_3 \mu_4})=\mathcal{S}(1,2,3,4)=1\,,\\
 \mathcal{S}(\eta^{\mu_1 \mu_3}\eta^{\mu_2 \mu_4})=\mathcal{S}(1,3,2,4)=-1\,.
\end{align}
Here it is important to employ the symmetric property of $\eta$ in order to sort w.r.t. to the order within each metric tensor.
That means:
\begin{align}
 \mathcal{S}(\eta^{\mu_2 \mu_1})=\mathcal{S}(\eta^{\mu_1 \mu_2})=\mathcal{S}(1,2)=1\,.
\end{align}
In the pictorial birdtracks representation one just has to count the number of crossings of metric tensor lines in a term in order to determine its sign.
If the count is even/odd the sign is ${+1}/{-1}$.

This in principle allows us to evaluate $T_{N}$ for arbitrary even $N$. 
However, in practical applications this is not very useful,
because the number of generated terms becomes too large to be treated in a fast way, 
even on modern computers.
However, we want to remark here, that up to now we have never used any information about the dimension of the vector representation $N_V$.
That means the given reduction is in fact independent of $N_V$.
The only dependence on $N_V$ is hidden inside the metric tensor $\eta$.

\subsection{Irreducible Representations and Dirac Bilinears}
In the following we will investigate the properties of the trace using the properties of irreducible representations.
This will allow us to evaluate traces with an odd number of Dirac matrices.
Assume we take the following contractions of a trace with $N=n_1+n_2$ Dirac matrices:
\begin{align}
 A_{n_1} \cdot T_{n_1+n_2}\cdot A_{n_2}=&A_{n_1}(\nu_1,\dots,\nu_{n_1};\mu_1,\dots,\mu_{n_1})T_{n_1+n_2}^{\mu_1\dots \mu_{n_1}\rho_{n_2}\dots\rho_1} A_{n_2}(\rho_1,\dots,\rho_{n_2};\sigma_1,\dots,\sigma_{n_2})\,.
\end{align}
Due to the property of the irreducible representation the trace can only be non-vanishing if the irreps $\hat{V}_{n_1}$ and $\hat{V}_{n_2}$ are equal.
In literature this is often called Schur's lemma.
When working with even or continuous $N_V$ we must have $n_1=n_2$.
We also know that the total result must be proportional to $A_{n_1}$ in this case, otherwise we would not stay in the sub space of the given irrep.
Up to a normalization constant -- which is given by the dimension of the spin space times $n_1!$ -- we can write for even or continuous $N_V$:
\begin{align}
 A_{n_1} \cdot T_{n_1+n_2}\cdot A_{n_2}= N_S n_1! A_{n_1} \delta_{n_1 n_2}\,.\label{EQ:IrrepOrthogonality}
\end{align}
For odd $N_V$ we get further contributions on the r.h.s.\ of the above equation whenever $n_1\neq n_2$ but $N_{\hat{V}_{n_1}}=N_{\hat{V}_{n_2}}$ holds.
This is the case whenever $n_1+n_2=N_V$.
As a special case we have for example $n_1=0$ and $n_2=N_V$:
\begin{align}
 A_{N_V} \cdot T_{N_V+0}\cdot A_{0}= -i^{(N_V-1)N_V/2} N_S \varepsilon_{N_V}\,.\label{EQ:AnTRA0}
\end{align}
And in the $n_1+n_2=N_V$ case the r.h.s.\ of corresponding equation is always proportional to $\varepsilon_{N_V}$.
The reason why this gives a contribution in odd dimensions only, and where the phase factor comes from, will be explained later.

When we define the Dirac bilinears
\begin{align}
 \Gamma^{\nu_1\dots\nu_{n}}_n=A_{n}(\nu_1,\dots,\nu_{n};\mu_1,\dots,\mu_{n}) \gamma^{\mu_1}\dots \gamma^{\mu_n}\,,
\end{align}
where the Dirac matrices on the right are multiplied with each other through the spinor space indices,
we can rewrite Eq.~(\ref{EQ:IrrepOrthogonality}) in terms of $\Gamma$'s:
\begin{align}
 A_{n_1} \cdot T_{n_1+n_2}\cdot A_{n_2}={\rm tr} \{\Gamma^{\mu_1\dots\mu_{n_1}}_{n_1} \Gamma^{\nu_1\dots\nu_{n_2}}_{n_2}\}= n_1!N_S \delta_{n_1 n_2}A_{n_1}(\mu_1,\dots,\mu_{n_1};\nu_{n_2},\dots,\nu_{1})\,.
\end{align}
That means the different  $\Gamma$'s are orthogonal to each other and they form a basis of matrices in spin space.
The basis allows one to decompose any product of Dirac matrices in terms of a linear combination of $\Gamma$'s:
\begin{align}
 \gamma^{\mu_1}\dots \gamma^{\mu_N}= \sum_{n}c^{\mu_1\dots \mu_N}_{n,\nu_1\dots\nu_n}\Gamma^{\nu_1\dots\nu_n}_{n}\,.\label{EQ:BilinearDecomposition}
\end{align}
The coefficients $c_n$ are in fact coefficient tensors built up by products of metric tensors.
In odd dimensions $N_V= 2m+1$ there may also appear the tensor $\varepsilon_{2m+1}$.
The sum runs over all values of $n=0,\dots,N_V$ when $N_V$ is even, 
$n=0,\dots,(N_V-1)/2$ when $N_V$ is odd,
and from zero to infinity for a non-integer $N_V$.
In the latter case we are thus dealing with an infinite dimensional algebra, because there are countable but infinitely many linearly independent Dirac bilinears. 
As an example we can reduce the product of two Dirac matrices into a sum of two terms:
\begin{align}
 \gamma^{\mu_1}\gamma^{\mu_2}=\tfrac{1}{2}\{\gamma^{\mu_1},\gamma^{\mu_2}\}+\tfrac{1}{2}[\gamma^{\mu_1},\gamma^{\mu_2}]=\eta^{\mu_1\mu_2}\Gamma_0+\Gamma^{\mu_1\mu_{2}}_{2}\,.
\end{align}
We have $\Gamma_0=1_{N_S\times N_S}$.
Further we note that any  $\Gamma_n$ is irreducible with respect to Eq.~(\ref{EQ:DA}),
because the fully anti-symmetric vector indices have no symmetric subsets.
The number of linearly independent Dirac bilinears is given by the sum of the dimensions of all independent irreps :
\begin{align}
 N_B=\sum\limits_{n=0}^{n_B-1} N_{\hat{V}_n}\,.
\end{align}
For $d=2$ ($n_B=3$) we find  $N_B=1+2+\underline{1}=4$.
For $d=3$ ($n_B=2$) we have $N_B=1+\underline{3}=4$.\\
For $d=4$  ($n_B=5$) we obtain $N_B=1+4+\underline{6}+4+1=16$.
The underlined number displays $N_A$ respectively.
In the literature the $\varepsilon_{N_V}$ tensors are absorbed into the $\Gamma_n$'s.
For example one can choose $1_{N_S \times N_S}$, $\gamma^{\mu}, \sigma^{\mu\nu}=\Gamma_2^{\mu \nu}, \gamma_5 \gamma^{\mu}$ and $\gamma_5$ to be the 16 Dirac bilinears in four dimensions.
This has the advantage that the coefficient tensors have a smaller rank.

We can define special scalar bilinears for any integer dimension $N_V=n$ via:
\begin{align}
 \gamma_{n+1}= i^{n(n-1)/2}[\varepsilon_{n}]_{\mu_1\dots\mu_{n}}\Gamma_{n}^{\mu_1\dots\mu_{n}}=i^{n(n-1)/2}\gamma^{1}\dots \gamma^{n}\,.\label{EQ:DEFGAMMANV+1}
\end{align}
With the chosen normalization factor (and euclidean metric $\eta$) one can check using Eq.~(\ref{EQ:DA}) that $\gamma_{n+1}\cdot\gamma_{n+1}=1_{N_S\times N_S}$.
We further see that we have:
\begin{align}
 [\gamma_{2n},\gamma^{\mu}]=&0\,,&\{\gamma_{2n+1},\gamma^{\mu}\}=&0\,.
\end{align}
That means for odd $N_V$ we have a scalar $\gamma_{N_V+1}$ that commutes with all other $\gamma^\mu$. 
Whereas in even $N_V$ we have a pseudo scalar $\gamma_{N_V+1}$ that anti-commutes with all other $\gamma^\mu$. 

Because $\gamma_{2n}$ commutes with all $\gamma^{\mu}$ it must be proportional to $1_{N_S\times N_S}$.
Since its square is $1_{N_S\times N_S}$ we can only have $\gamma_{2n}=\pm 1_{N_S\times N_S}$.
That means ${\rm tr}\{\gamma_{2n}\}=\pm N_S$.
Because we want to follow the convention chosen in the explicit representation of $\gamma^\mu$ for $N_V=3$ given by the Pauli matrices we have to stick with the plus sign.
This is in fact the reason for the phase factor in Eq.~(\ref{EQ:AnTRA0}). 
Thus contracting Eq.~(\ref{EQ:AnTRA0}) with $\varepsilon_{2n}$ and using the definition of $\gamma_{2n}$ yields its r.h.s.\ .

We already found a case in odd dimensions where a trace of an odd number of Dirac matrices yields a non-vanishing result.
Are there more cases for lower $n<N_V$?
The short answer is: No, because one has to have sufficiently many vector indices to saturate the rank $N_V$ of  $\varepsilon_{N_V}$.
And it is the only invariant tensor with an odd number of vector indices.
To prove this intuitive statement we can calculate for integer $N$:
\begin{align}
N_V {\rm tr}\{\Gamma_{N}^{\mu_1\dots\mu_{N}}\}=& {\rm tr}\{\gamma^{\nu}\gamma_{\nu}\Gamma_{N}^{\mu_1\dots\mu_{N}}\}\nonumber\\
=&2N{\rm tr}\{\Gamma_{N}^{\mu_1\dots\mu_{N}}\}+(-1)^N{\rm tr}\{\gamma^{\nu}\Gamma_{N}^{\mu_1\dots\mu_{N}}\gamma_{\nu}\}\nonumber\\
=&2N{\rm tr}\{\Gamma_{N}^{\mu_1\dots\mu_{N}}\}+(-1)^NN_V{\rm tr}\{\Gamma_{N}^{\mu_1\dots\mu_{N}}\}\,.
\end{align}
Here we used Eq.~(\ref{EQ:AntiCommutingThroughgammas}) in order to anti-commute $\gamma_{\nu}$ through $\Gamma_{N}^{\mu_1\dots\mu_{N}}$.
Splitting the above equation into even $N=E>0$ and odd $N=O$ case yields
\begin{align}
 (O-N_V){\rm tr}\{\Gamma_{O}^{\mu_1\dots\mu_{O}}\}=&0\,,& {\rm tr}\{\Gamma_{E}^{\mu_1\dots\mu_{E}}\}=&0\,.\label{EQ:TROdRestrictions}
\end{align}
So only for $N_V=O$ the trace is allowed to have a non-vanishing result.
And this is exactly the case of ${\rm tr}\{\gamma_{2n}\}$.
But that means a trace of an odd number $n$ of Dirac matrices for odd $N_V$ is only non-zero for $n\geq N_V$.
However, the source of such a non-vanishing result is only given by the coefficient tensor of the $\Gamma_{N_V}^{\mu_1\dots\mu_{N_V}}$ bilinear
in the bilinear decomposition.
In this case the coefficient tensor $c_{N_V-m}$ is not linearly independent to $c_{m}$ for $m\in\{0,1,\dots,N_V\}$.
For $m=0$ we have for example:
\begin{align}
 c^{\nu_1\dots \nu_N}_{N_V,\mu_1 \dots \mu_{N_V}}i^{N_V(N_V-1)/2}\varepsilon_{N_V}^{\mu_1 \dots \mu_{N_V}}=c_0^{\nu_1\dots \nu_{N}}\,.
\end{align}
Which is why for odd $N_V$ the sum of Eq.~(\ref{EQ:BilinearDecomposition}) runs only from zero to $(N_V-1)/2$.
For example for $N_V=3$ we see that $A_2$ projects onto the same irrep like $A_1$.
We can immediately see this, because the generators of $SO(3)$ are given by $T^\rho_{\mu\nu}\sim\epsilon_{\mu\nu\rho}$,
which means for $SO(3)$ the vector index $\mu$ agrees with the adjoint index $a$ and thus $N_A=N_V$.

Due to the discussed property of $\gamma_{2n}$ one can always reduce a trace with an odd number of Dirac matrices to a trace with an even number.
Because we have:
\begin{align}
 T_{2n+1}^{\mu_1\dots\mu_{2n+1}}=&T_{2n+1+N_V}^{N_V+1 \mu_1\dots\mu_{2n+1}}={\rm tr}\{\gamma_{N_V+1}\gamma^{\mu_1} \dots\gamma^{\mu_{2n+1}}\}\,.\label{EQ:OddToEvenTrace}
\end{align}

For even $N_V$ the pseudo scalar $\gamma_{N_V+1}$ allows us to split the spinor space into distinct sub-spaces,
because one can construct the projectors:
\begin{align}
 P^{\pm}=\frac{1_{N_S\times N_S}\pm\gamma_{2n+1}}{2}\,,
\end{align}
which project onto the left- and right-handed spinor representation.

In order to investigate the possible results for traces with a single $\gamma_{2n+1}$ we calculate:
\begin{align}
  N_V{\rm tr}\{\gamma_{2n+1}\Gamma_{N}^{\mu_1\dots\mu_{N}}\}=&{\rm tr}\{\gamma_{2n+1}\gamma^{\nu}\gamma_{\nu}\Gamma_{N}^{\mu_1\dots\mu_{N}}\}\nonumber\\
  =& 2N{\rm tr}\{\gamma_{2n+1}\Gamma_{N}^{\mu_1\dots\mu_{N}}\}-(-1)^N N_V{\rm tr}\{\gamma_{2n+1}\Gamma_{N}^{\mu_1\dots\mu_{N}}\}\,.\label{EQ:gammaNV+1ForNVEven}
\end{align}
In the last line we have again used Eq.~(\ref{EQ:AntiCommutingThroughgammas}) to anti-commute $\gamma_{\nu}$ through $\Gamma_{N}^{\mu_1\dots\mu_{N}}$.
Further we used the cyclicity of the trace and the anti-commutativity of $\gamma_{2n+1}$ (which gives one additional minus sign).
Splitting the even $N=E$ and odd $N=O$ case we get the following constraints:
\begin{align}
 (N_V-E){\rm tr}\{\gamma_{2n+1}\Gamma_{E}^{\mu_1\dots\mu_{E}}\}=&0\,,&{\rm tr}\{\gamma_{2n+1}\Gamma_{O}^{\mu_1\dots\mu_{O}}\}=&0\,.\label{EQ:TREdRestrictions}
\end{align}
Thus we can only have a non-vanishing result of the trace of a single $\gamma_{2n+1}$ when we have at least $N_V=E=2n$ many other Dirac matrices in it.
As important special case we have:
\begin{align}
 {\rm tr}\{\gamma_{2n+1}\}=0\,.
\end{align}
Which is why Eq.~(\ref{EQ:AnTRA0}) does only apply for odd $N_V$.
Or in other words in odd dimensions the anti-symmetrizer $A_{N_V}$ projects onto the singlet/scalar irrep.
In even dimensions the anti-symmetrizer $A_{N_V}$ projects on a pseudo singlet/scalar irrep, which is not equivalent to the singlet/scalar irrep.

However, when one deforms the dimension away from integer values for example into the real numbers -- like it is done during any Dimensional Regularization -- 
one cannot obtain any fully anti-symmetric $\varepsilon_{N_V}$-tensor, 
if one considers $\gamma_{N_V+1}$ to be naively (anti\mbox{-})\-~commuting when the traces are cyclic.
Because the trace can only be non-zero when the prefactor is zero.
That means following a naive (anti\mbox{-})\-~commuting prescription for $\gamma_{N_V+1}$ in generic $N_V\notin\mathbb{N}$ dimensions,
forces one to set all traces with a single $\gamma_{N_V+1}$ in it to zero, if one obeys the Clifford algebra.
In fact this is a major obstacle for the continuation of for example $N_V$ dimensional amplitudes into $d$ dimensions,
because there is no smooth limit of the abstract $d$-dimensional Clifford algebra to an integer dimensional one, 
that does preserve all intrinsically integer dimensional contributions.

\section{Dimensional Continuation in Presence of $\varepsilon_{N_V}$}\label{SEC:DC}
Eq.~(\ref{EQ:TROdRestrictions}) and Eq.~(\ref{EQ:TREdRestrictions}) indicate that one cannot 
carry out the Clifford algebra in non-integer $d=N_V$ dimensions AND retain any $\varepsilon_{N_V}$ contributions if one 
believes that cyclicity of the trace and naive-{(anti\mbox{-})}~commutativity of $\gamma_{N_V+1}$ is required in order to retain the symmetries of the original $\hat{N}_V$ integer dimensional traces.

We will call $\hat{N}_V$ target dimension, because we are interested in limits of the form $N_V\rightarrow \hat{N}_V$.
For example for $\hat{N}_V=4$ one often uses  $N_V=4-2\epsilon$ where $\epsilon\rightarrow0$ is a small parameter and regulates divergent quantities in terms of inverse powers of $1/\epsilon$.

In order to have a regulator that respects the $\hat{N}_V$ dimensional symmetries (which is important for the preservation of Ward-identities) 
we must require that any traces that are equivalent due to cyclicity or the (anti-) commutation property of $\gamma_{N_V+1}$ in $\hat{N}_V$ dimensions must be continued in the same way to $N_V$ dimensions.
Otherwise we break the assertion that a symmetry preserving regulator has to continue $\hat{N}_V\rightarrow N_V$ in a uniform way.
This means $\gamma_{N_V+1}$ has to be treated to (anti\mbox{-})\-~commute with all other $\gamma^{\mu}$.
Further cyclic reordering of traces must not change the result.

At first sight the given requirements seem to be easily obeyed. 
One just uses the $\hat{N}_V$ dimensional Clifford algebra (where $\gamma_{N_V+1}$ (anti-)commutes) 
to reduce fermion traces in terms of $\hat{\eta}$ (the hat indicates that the dimension of $\eta$ is given by the integer $\hat{N}_V$) and $\varepsilon_{N_V}$.
Using a proper projection in terms of $\varepsilon_{N_V}$ allows us to eliminate it in favor of $A_{N_V}$ (See Eq.~(\ref{EQ:AnSplitTo2eps})).
Then one is free to trivially continue $\hat{\eta}\rightarrow \eta$.

This is what for example can be done and was done in order to evaluate the SM $\beta$-function at three-loop level for $\hat{N}_V=4$ ~\cite{Mihaila:2012fm,Mihaila:2012pz,Chetyrkin:2012rz}.
Meaning the traces can be evaluated using the \verb|trace4| algorithm implemented in the computer algebra program {\tt FORM}~\cite{Vermaseren:2000nd,Kuipers:2012rf,Ruijl:2017dtg}
when one declares all appearing vector indices and vectors to be non-integer dimensional.

We have even checked that the \verb|trace4| algorithm works for the chiral-XY Gross-Neveu-Yukawa model up to including four loops for $\hat{N}_V=4$.
That means in this case we obtain the same $Z$-factor results like when using a naive-anti-commuting $\gamma_5$ prescription previously employed in Ref.~\cite{Zerf:2017zqi}.

The function \verb|trace4| eliminates all trace internal contraction indices 
and the appearance of two equivalent vectors using the anti-commutation relation via Eq.~(\ref{EQ:AntiCommutingThroughgammas}).
In case it identifies four dimensional vector indices in the trace further $N_V=4$ dimensional reduction rules\footnote{What they are in detail can be read off the online manual~\cite{FORMONLINMANUAL}.} are applied.
After that any trace $T_N$ contains $N$ different indices and/or vectors and the  $N_V=4$ dimensional rule\footnote{$\tilde{\epsilon}_4$ is defined like in {\tt FORM} and agrees up to a phase with $\epsilon_4$.}:
\begin{align}
 \gamma_{\mu_1}\gamma_{\mu_2}\gamma_{\mu_3}&= \gamma_{5}\gamma_{\nu}[\tilde{\epsilon}_4]_{\mu_1\mu_2\mu_3\nu}+\eta_{\mu_1\mu_2}\gamma_{\mu_3}-\eta_{\mu_1\mu_3}\gamma_{\mu_2}+\eta_{\mu_2\mu_3}\gamma_{\mu_1}\,, \label{EQ:3GammaDecompoIn4D}
\end{align}
is used to further reduce them. 
Eq.~(\ref{EQ:3GammaDecompoIn4D}) resembles the four dimensional bilinear decomposition of a product of three Dirac matrices.

However, the results obtained by \verb|trace4| for traces appearing in a four-loop SM $Z$-factor calculation turn out to be not suitable for a proper dimensional continuation in the general case.

To proceed we first have to understand the difference between traces reduced assuming a continuous and integer dimensional algebra.
In a first step we do this for traces that do not contain intrinsically $\hat{N}_V$ dimensional tensors.
Further, for simplicity we use $\hat{N}_V=2$, but the obtained insight generalizes to any $\hat{N}_V$.\\
Calculating the trace $T_6^{p q r s t u}$ using Eq.~(\ref{EQ:T6}) in $N_V$ dimensions yields in total $15$ terms.
Taking the limit $T_6^{p q r s t u}\big|_{N_V\rightarrow 2}$ by explicitly plugging in two dimensional vectors for $p,q,r,s,t$ and $u$ 
as well as replacing $\eta\rightarrow \hat{\eta}$ with $\hat{\eta}={\rm diag}( 1, 1)$
yields $32$ terms. 
One of them is for example $+p_1 q_1 r_1 s_1 t_1 u_1$.
These $32$ terms remain unchanged when dropping a collection of six of the original $15$ terms. 
For example the six terms contained in
\begin{align}
 6 A_3(p,q,r;u,t,s)=\left[(p\cdot u)(q\cdot t) (r\cdot s)-(p\cdot t)(q\cdot u) (r\cdot s) +\dots\right]\,,\label{EQ:A3EVAN}
\end{align}
do the job (here and in the following we use $p \cdot q= p \cdot \eta\cdot q$).
That means:
\begin{align}
 T_6^{p q r s t u}\Big|_{N_V\rightarrow 2}=&\left(T_6^{p q r s t u}- 6 N_S A_3(p,q,r;u,t,s)\right)\big|_{N_V\rightarrow 2}\,.
\end{align}
Clearly we must have:
\begin{align}
 A_3(p,q,r;u,t,s)\big|_{N_V\rightarrow 2}=& 0\,.
\end{align}
The reason for this identity is that one cannot anti-symmetrize more than two indices in two dimensions,
because any third index carries either the value of the first or second one and is thus linearly dependent.
For general $\hat{N}_V$ we have:
\begin{align}
 A_{N}(\dots;\dots)\big|_{N_V\rightarrow \hat{N}_V}=& 0\,,&\forall N>& \hat{N}_V\,.
\end{align}
In the following we call tensors or contributions that obey the above equation evanescent tensors or contributions.
And as soon as we do not eliminate appearing evanescent contributions which are exactly zero for $N_V=\hat{N}_V$
we risk breaking the required uniformity of the continuation to non-integer $N_V$ dimensions.
In the language of the authors of Ref.~\cite{Breitenlohner:1977hr} 
an expression that has gone through a complete elimination of all appearing evanescent structures is said to be in ``normal form''.
Like these authors already pointed out, there is (unfortunately) not a unique normal form.
Because one has the choice which tensor contraction is eliminated in favor of the others.
Back to our example with $A_3$ one can for example eliminate
\begin{align}
 (p\cdot u)(q\cdot t) (r\cdot s) \rightarrow -6 A_3(p,q,r;u,t,s)+(p\cdot u)(q\cdot t) (r\cdot s) \,,
\end{align}
the first terms on the l.h.s of Eq.~(\ref{EQ:A3EVAN}) in favor of the remaining ones.
Thus the first term will not appear in the expression anymore.
But one could also choose to eliminate the second term and so on.
Of course any consistent elimination requires that one always eliminates exactly the same conventionally chosen term.
But before one is able to eliminate evanescent contributions one has to identify them in a complete way.
We perform this task in the next section.

\section{Maximal Decomposition of Spintraces}\label{SEC:MAXDEC}
In order to avoid the appearance of evanescent structures during the evaluation of spin traces $T_N$,
one might have the idea to use the iterative decomposition of products of Dirac bilinears
over the finite set of $n_B$ $\Gamma_i$'s available in $\hat{N}_V$ dimensions (employing Eq.~(\ref{EQ:BilinearDecomposition})):
\begin{align}
 \Gamma_{n_1} \Gamma_{n_2} = \sum\limits_{i=0}^{n_B-1} c^{n_1 n_2}_i \Gamma_{i}\,.
\end{align}
Here we suppressed all vector indices and assume a multiplication in spin space on the l.h.s. of the equation.
However, example evaluations for $\hat{N}_V=4$ have shown, that doing so, the coefficient tensor of $\Gamma_0$ emerging from the very last decomposition,
contains an evanescent contribution in the result of $T_N$ starting from $N=10$,
because the obtained result was given by a sum of $9!!=945$ terms and exactly agreed with the result of a generic $N_V$ dimensional reduction
and therefore no terms have been eliminated.
It is clear that in the generic $N_V$ dimensional result one can group up $5!$ terms to form an $\hat{N}_V=4$ evanescent anti-symmetrizer $A_5$.

For the calculation of $T_{2N}$ for $N\leq  \hat{N}_V$ however, the finite basis decomposition yields evanescent free results usable for dimensional continuation.
Here the decomposition keeps $\gamma_{N_V+1}$ (anti\mbox{-})~commuting, because it does not allow for a decomposition of the type 
\begin{align}
 [\epsilon_{N_V}]_{\nu_1 \dots \nu_{N_V}}\Gamma_{N_V}^{\nu_1 \dots \nu_{N_V}}\Gamma_{1}^{\mu}\sim [\epsilon_{N_V}]_{\nu_1 \dots \nu_{N_V}} \Gamma_{N_V+1}^{\nu_1 \dots \nu_{N_V} \mu}+\dots\,,
\end{align}
The ellipsis indicates a non-evanescent $\Gamma_{N_V-1}$ bilinear contribution.
Because such a decomposition would immediately run into an evanescent contribution just from merging the product $\gamma_{\hat{N}_V+1}\gamma^{\mu}$.

\begin{figure}
\begin{tabular}{c}
  \includegraphics{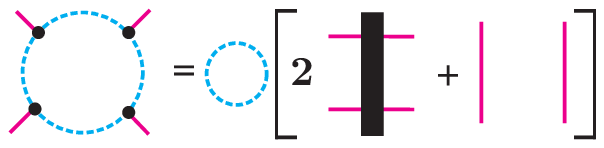}\\
  \includegraphics{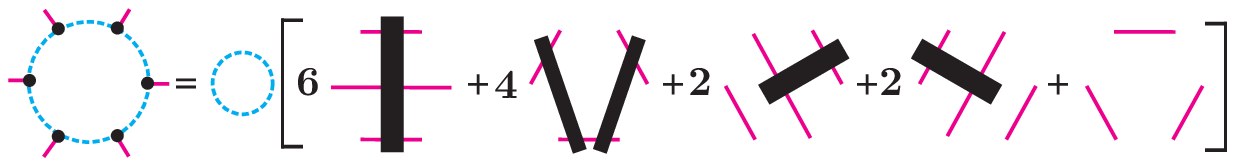}\\
  \includegraphics{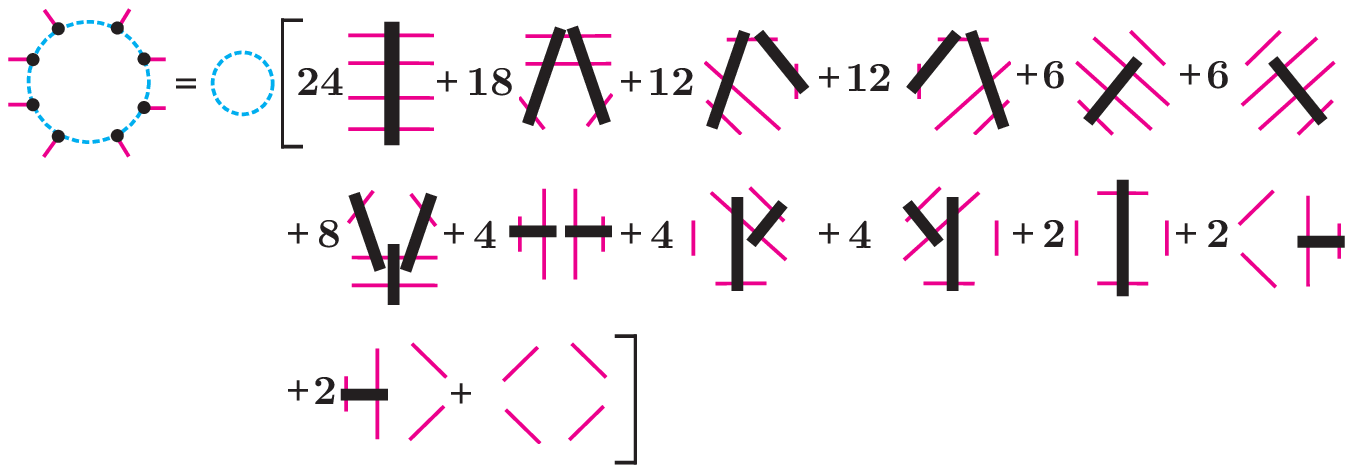}
\end{tabular}
\caption{\label{FG:DECOMPN}Results for decompositions of $T_N$ for the first 3 non-trivial $N$ in the pictorial birdtrack notation already used in Fig.~\ref{FIG:TrN}.
A black bar connected with $2n$ solid (magenta) lines correspond to an anti-symmetrizer $A_n$.}
\end{figure}

For $N> \hat{N}_V$ we suggest to proceed as follows in order to gain control over the evanescent contributions.
We generate the full result $T_N$ and perform a maximal decomposition in the sense that we pack as many terms as possible into anti-symmetrizers starting with the one of maximal rank $N/2$.
For example we can choose:
\begin{align}
T_{4}^{\mu_1\mu_2\mu_3\mu_4}=& N_S\left[\eta^{\mu_1 \mu_4}\eta^{\mu_2 \mu_3} -\eta^{\mu_1 \mu_3}\eta^{\mu_2 \mu_4}+\eta^{\mu_1 \mu_2}\eta^{\mu_3 \mu_4}\right]\nonumber\\
=&N_S \left[2A_2(\mu_1, \mu_2;\mu_4, \mu_3)+A_1(\mu_1; \mu_2)A_1(\mu_3; \mu_4)\right]\,,\label{EQ:T4Dec}\\ 
T_{6}^{\mu_1\mu_2\mu_3\mu_4\mu_5\mu_6}=&N_S\big[~6A_3(\mu_1, \mu_2, \mu_3;\mu_6, \mu_5, \mu_4)+4A_2(\mu_1,\mu_2;\nu,\mu_3) A_2(\nu,\mu_4;\mu_6,\mu_5)\nonumber\\
&+2A_2(\mu_3,\mu_4;\mu_6,\mu_5)A_1(\mu_1;\mu_2)+2A_2(\mu_1,\mu_2;\mu_4,\mu_3)A_1(\mu_5;\mu_6)\nonumber\\
&+A_1(\mu_1;\mu_2)A_1(\mu_3;\mu_4)A_1(\mu_5;\mu_6)\big]\,,\label{EQ:T6Dec}\\ 
T_{8}^{\mu_1\mu_2\mu_3\mu_4\mu_5\mu_6\mu_7\mu_8}=&N_S\big[~24A_4(\mu_1, \mu_2, \mu_3,\mu_4;\mu_8, \mu_7, \mu_6, \mu_5)\nonumber\\
    &+18 A_3(\mu_2, \mu_3, \mu_4; \mu_1, \nu_1, \nu_2) A_3(\mu_8, \nu_1, \nu_2; \mu_7, \mu_6, \mu_5)\nonumber\\
    &+12 A_2(\mu_5, \mu_6; \nu_1, \mu_7) A_3(\mu_4, \mu_3, \mu_2; \nu_1, \mu_8, \mu_1)\nonumber\\
    &+12 A_2(\mu_4, \mu_3; \nu_1, \mu_2) A_3(\nu_1, \mu_1, \mu_8; \mu_5, \mu_6, \mu_7)\nonumber\\
    &+6 A_1(\mu_3; \mu_4) A_3(\mu_2, \mu_1, \mu_8; \mu_5, \mu_6, \mu_7)\nonumber\\
    &+6 A_1(\mu_5; \mu_6) A_3(\mu_2, \mu_3, \mu_4; \mu_1, \mu_8, \mu_7)\nonumber\\
    &+8 A_2(\mu_2, \mu_3; \nu_1, \mu_4) A_2(\mu_1, \nu_1; \mu_8, \nu_2) A_2(\nu_2, \mu_5; \mu_7, \mu_6)\nonumber\\
    &+4 A_2(\mu_2, \mu_1; \mu_3, \mu_4) A_2(\mu_5, \mu_6; \mu_8, \mu_7)\nonumber\\
    &+4 A_1(\mu_2; \mu_3) A_2(\mu_4, \mu_1; \nu_1, \mu_8) A_2(\nu_1, \mu_5; \mu_7, \mu_6)\nonumber\\
    &+4 A_1(\mu_6; \mu_7) A_2(\nu_1, \mu_1; \mu_5, \mu_8) A_2(\mu_2, \mu_3; \nu_1, \mu_4)\nonumber\\
    &+2 A_1(\mu_2; \mu_3) A_1(\mu_6; \mu_7) A_2(\mu_1, \mu_4; \mu_8, \mu_5)\nonumber\\
    &+2 A_1(\mu_3; \mu_4) A_1(\mu_1; \mu_2) A_2(\mu_5, \mu_6; \mu_8, \mu_7)\nonumber\\
    &+2 A_1(\mu_5; \mu_6) A_1(\mu_7; \mu_8) A_2(\mu_1, \mu_2; \mu_4, \mu_3)\nonumber\\
    &+A_1(\mu_1; \mu_2) A_1(\mu_3, \mu_4) A_1(\mu_5, \mu_6) A_1(\mu_7, \mu_8)\big]\,.\label{EQ:T8Dec}
\end{align}
A pictorial representation of the selected solutions for the first three decompositions is shown in Fig.~\ref{FG:DECOMPN}.
The ordering of terms in the figure is in agreement with the one in the equation.
$\mu_1$ is chosen to be the index in the lower left corner of each graph.

The decomposition is nothing but a special partition of terms made up by products of metric tensors that appear in the trace result.
Because all ways of finding the maximal rank $N/2$ anti-symmetrizer in a trace result share the same ``inversion contraction'' term\footnote{for $T_4$/$T_6$ it is given by the second/8th term shown in Fig.~\ref{FIG:TrN}}, which stays invariant under every cyclic rotation,
one can only have one subset containing a maximal anti-symmetrizer in the decomposition.
It is further clear, that one can build only finitely many anti-symmetrizer of rank $N/2-1$ from the remaining terms of a trace and so on.

We call the displayed decomposition maximal, because the number of subsets represented by a product of anti-symmetrizers ($\eta=A_1$) is minimal.
We do not attempt to prove this statement here, but just note that the minimal number of subsets is given by the number of planar diagrams appearing in the pictorial birdtracks representation of the traces in Fig.~\ref{FIG:TrN}.
This is so, because a product of anti-symmetrizers can only contain a single planar contribution.
It can be obtained by removing all black bars indicating anti-symmetrizers.
The number  $P$ of planar birdtracks graphs in dependence of the even number $N$ of $\gamma$ matrices is given by:
\begin{align}
 P(N)=\frac{N!}{\left(\tfrac{1}{2}N\right)!\left(\tfrac{1}{2}N+1\right)!}\,.
\end{align}
The number of planar graphs grows moderately fast with $N$ as the following table shows.
\begin{center}
 \begin{tabular}{c|c}
 $N$& $P(N)$\\
 \hline
 $2$ & $1$\\
  $4$ & $2$\\
  $6$ & $5$\\
  $8$ & $14$\\
  $10$ & $42$\\
  $12$ & $132$
\end{tabular}
\end{center}
Another property of a maximal decomposition is that it always contains a single subset containing only one term, a planar product of metric tensors.
Of course there is in general not a unique solution for a maximal decomposition and in general it would be interesting to find all possible solutions, their properties and relations.

The displayed solutions have an additional, very important property which can be immediately discovered in the birdtracks graphs.
They are symmetric with respect to mirror operations along the vertical axis.
That means either the drawn subset transforms into itself, because it is mirror symmetric, or a pair of subsets transforms into each other.

Besides the identification of evanescent contributions another benefit of the maximal decomposition is that one gets a handle on the symmetries of the full trace results.
That means most of the time it is possible to eliminate complete subsets at once, because the external tensors are not sufficiently anti-symmetric to sustain higher rank anti-symmetrizers.
Such an elimination can be carried out at sub-set level and does not require an explicit, time and term heavy expansion of the contained $A_n$'s.
For example one can easily calculate the trace $T_8^{p p p p p p p p}$. 
Because it is  not possible to saturate any $A_n$ for $n>1$ we are immediately left with the last term of the decomposition in Fig.~\ref{FG:DECOMPN}.
The result is thus given by $T_8^{p p p p p p p p}=N_S(p\cdot p)^4$.

\section{Traces for Dimensional Continuation}\label{SEC:TRDC}
In this section we are going to determine how to evaluate traces such that they are suitable for a dimensional continuation in more detail.
We do this by carrying out simple example evaluations in order to find a suitable reduction prescription.

First we evaluate the following trace for the explicit target dimension $\hat{N}_V=2$:
\begin{align}
 T_{t_1}={\rm tr}\{\gamma_{3}\gamma_{3}\}=-\tfrac{1}{4}[\epsilon_{2}]_{\mu_1\mu_2}[\epsilon_{2}]_{\mu_3\mu_4}{\rm tr}\{\gamma^{\mu_1}\gamma^{\mu_2}\gamma^{\mu_3}\gamma^{\mu_4}\}\nonumber\\
 =-\tfrac{1}{4}[\epsilon_{2}]_{\mu_1\mu_2}[\epsilon_{2}]_{\mu_3\mu_4} 2 N_S A_2(\mu_1,\mu_2;\mu_4,\mu_3)\,.
\end{align}
Here we immediately dropped the last term of the decomposition of $T_4$ Eq.~(\ref{EQ:T4Dec}),
because contracting an anti-symmetric tensor with a symmetric one yields zero.
Using Eq.~(\ref{EQ:AnSplitTo2eps}) to eliminate the $\epsilon_2$'s yields:
\begin{align}
 T_{t_1}=-N_S A_2(\mu_1,\mu_2;\mu_3,\mu_4) A_2(\mu_1,\mu_2;\mu_4,\mu_3)\,.
\end{align}
Using the projector property $A_2\cdot A_2=A_2$ allows us to write:
\begin{align}
 T_{t_1}=-N_S A_2(\mu_4,\mu_3;\mu_3,\mu_4)=N_S A_2(\mu_4,\mu_3;\mu_4,\mu_3)=N_S N_{\hat{V}_2}=N_S\tfrac{1}{2}N_V(N_V-1) \,.
\end{align}
Only for $N_V=\hat{N}_V=2$ we obtain the expected result of $T_{t_1}=N_S$ for an anti-commuting $\gamma_3$.
Thus we need to require that the indices carried by $\epsilon_{N_V}$ tensors have to be $\hat{N}_V$ dimensional,
if we want to keep the symmetries of the $\hat{N}_V$ dimensional space.
Thus from now on we indicate intrinsically $\hat{N}_V$ dimensional indices with a hat like $\hat{\mu}$
in order to distinguish them from continuous dimensional indices, which are carried by trace external tensors like momenta or vector indices carried by propagating vector bosons.

We then compute:
\begin{align}
 T_{t_2}={\rm tr}\{\gamma_{3}\slashed{p}\gamma_{3}\slashed{q}\}=-\tfrac{1}{4}p^{\nu_1}q^{\nu_2}[\epsilon_{2}]_{\hat{\mu}_1\hat{\mu}_2}[\epsilon_{2}]_{\hat{\mu}_3\hat{\mu}_4}{\rm tr}\{\gamma^{\hat{\mu}_1}\gamma^{\hat{\mu}_2}\gamma^{\nu_1}\gamma^{\hat{\mu}_3}\gamma^{\hat{\mu}_4}\gamma^{\nu_2}\}\,.
\end{align}
Plugging in  the maximal decomposition for $T_6$ (Eq.~\ref{EQ:T6Dec}) and dropping all terms that yield zero due to symmetry, yields:
\begin{align}
  T_{t_2}&=-\tfrac{1}{4}p^{\nu_1}q^{\nu_2}[\epsilon_{2}]_{\hat{\mu}_1\hat{\mu}_2}[\epsilon_{2}]_{\hat{\mu}_3\hat{\mu}_4}N_S\big[6 A_3(\hat{\mu}_1,\hat{\mu}_2,\nu_1;\nu_2,\hat{\mu}_4,\hat{\mu}_3)\nonumber\\
  &+4A_2(\hat{\mu}_1,\hat{\mu}_2;\overset{?}{\rho},\nu_1) A_2(\overset{?}{\rho},\hat{\mu}_3;\nu_2,\hat{\mu}_4)
   +2A_2(\hat{\mu}_1,\hat{\mu}_2,\hat{\mu}_3,\nu_1)\eta^{\hat{\mu}_4\nu_2}\big]\,.
\end{align}
The question mark above the internal symmation index $\rho$ indicates that up to now it is not clear if it should wear a hat.
Using Eq.~(\ref{EQ:AnSplitTo2eps}) to eliminate the $\epsilon_2$'s and contracting the obtained $A_2$ yields:
\begin{align}
   T_{t_2}&=+\tfrac{1}{2}p^{\nu_1}q^{\nu_2}N_S\big[6 A_3(\nu_1,\hat{\mu}_1,\hat{\mu}_2;\nu_2,\hat{\mu}_1,\hat{\mu}_2)
  +4A_2(\hat{\mu}_1,\hat{\mu}_2;\nu_1,\overset{?}{\rho}) A_2(\overset{?}{\rho},\hat{\mu}_1;\nu_2,\hat{\mu}_2)\nonumber\\
   &+2A_2(\hat{\mu}_1,\hat{\mu}_2,\nu_1,\hat{\mu}_1)\eta^{\hat{\mu}_2\nu_2}\big]\,.
\end{align}
Before we can simplify this expression further we have to define the relation between hatted (integer) and unhatted (continuous) dimensional indices.
There are two possible prescriptions:
\begin{enumerate}
 \item{The index $\mu$ can only take a sub-set of values of the index $\hat{\mu}$: $\mu \subseteq\hat{\mu}$.
       That means the hatted index range is bigger than the one without hat:
       \begin{align}
         \eta_{\mu \hat{\nu}}&=\eta_{\hat{\mu} \nu}=\eta_{\mu \nu}\,,\label{EQ:LooseHatPrescription1}\\
         \hat{\eta}_{\mu \hat{\nu}}&=\hat{\eta}_{\hat{\mu} \nu}=\eta_{\mu \nu}\,,\label{EQ:LooseHatPrescription2}\\
        \hat{\eta}\cdot\eta&=\eta\,.\label{EQ:LooseHatPrescription3}
       \end{align}
}
 \item{The index $\hat{\mu}$ can only take a sub-set of values of the index $\mu$: $\hat{\mu} \subseteq \mu$.
       That means the hatted index range is smaller than the one without hat:
       \begin{align}
         \eta_{\mu \hat{\nu}}&=\eta_{\hat{\mu} \nu}=\eta_{\hat{\mu} \hat{\nu}}=\hat{\eta}_{\hat{\mu} \hat{\nu}}=\hat{\eta}_{\mu \nu}\,,\\
        \hat{\eta}\cdot\eta&=\hat{\eta}\,.
       \end{align}
}
\end{enumerate}
In the first case the hats are ``loose'' and any metric tensor without hat hitting a metric with hat will make the hat drop off.
In the second case the hats are ``sticky'' and any metric tensor with a hat will never lose it.

Evaluating $T_{t_2}$ for the first case with ``loose'' hats immediately kills the evanescent contribution of $A_3$:
\begin{align}
 6 A_3(\nu_1,\hat{\mu}_1,\hat{\mu}_2;\nu_2,\hat{\mu}_1,\hat{\mu}_2) =\eta_{\nu_1 \nu_2}(\hat{N}_V-1)(\hat{N}_V-2)=0\,.
\end{align}
The second term inside the bracket of $T_{t_2}$ yields:
\begin{align}
4A_2(\hat{\mu}_1,\hat{\mu}_2;\nu_1,\overset{?}{\rho}) A_2(\overset{?}{\rho},\hat{\mu}_1;\nu_2,\hat{\mu}_2) =2A_2(\overset{?}{\rho},\nu_1;\nu_2,\overset{?}{\rho})=-\eta_{\nu_1 \nu_2}(\overset{?}{N}_V-1)\,.
\end{align}
And the last term in the bracket yields
\begin{align}
 2A_2(\hat{\mu}_1,\nu_2,\nu_1,\hat{\mu}_1) =-\eta_{\nu_1 \nu_2}(\hat{N}_V-1)\,,
\end{align}
That means, for lose hats we get:
\begin{align}
 T_{t_2}&=-N_S (p\cdot q) \big[\tfrac{1}{2}(\hat{N}_V+\overset{?}{N}_V)-1\big]\,.
\end{align}
This result only agrees with the intrinsically $\hat{N}_V=2$ dimensional one when we allow the internal contraction index $\rho$ -- appearing in the decomposition -- to wear a hat: $\overset{?}{\rho}\rightarrow \hat{\rho}$:
\begin{align}
 T_{t_2}&=-N_S (p\cdot q)\,.
\end{align}
Evaluating $T_{t_2}$ for the second case with ``sticky'' hats calling it $\hat{T}_{t_2}$ does NOT kill the evanescent contribution of $A_3$:
\begin{align}
 6 A_3(\nu_1,\hat{\mu}_1,\hat{\mu}_2;\nu_2,\hat{\mu}_1,\hat{\mu}_2)& =\eta_{\nu_1 \nu_2}\hat{N}_V(\hat{N}_V-1)-2\hat{\eta}_{\nu_1 \nu_2}(\hat{N}_V-1)\,.
\end{align}
The second term inside the bracket of $T_{t_2}$ yields:
\begin{align}
4A_2(\hat{\mu}_1,\hat{\mu}_2;\nu_1,\overset{?}{\rho}) A_2(\overset{?}{\rho},\hat{\mu}_1;\nu_2,\hat{\mu}_2)& =2A_2(\hat{\mu}_2,\hat{\nu}_1;\nu_2,\hat{\mu}_2)=-\hat{\eta}_{\nu_1\nu_2}(\hat{N}_V-1)\,.
\end{align}
Note that in this case the result is independent of the presence of a hat on the internal contraction index $\rho$, however we suggest to not give a hat to it in general.
For the last term inside the bracket we obtain:
\begin{align}
 2A_2(\hat{\mu}_1,\hat{\nu}_2,\nu_1,\hat{\mu}_1)&=-\hat{\eta}_{\nu_1\nu_2}(\hat{N}_V-1)\,.
\end{align}
We thus obtain:
\begin{align}
 \hat{T}_{t_2}&=+\tfrac{1}{2}N_S (\hat{N}_V-1)\big[ (p\cdot q) \hat{N}_V-4(\hat{p}\cdot \hat{q})\big]\nonumber\\
        &=N_S \big[ (p\cdot q)-2(\hat{p}\cdot \hat{q})\big]\,.
\end{align}
For $ (p\cdot q)\rightarrow(\hat{p}\cdot \hat{q})$ this result also reproduces the $\hat{N}_V$ dimensional result.
However, it shows that $\gamma_3$ does not anti-commute with all the $\gamma^{\mu}$ in continuous $N_V$ dimensions
and thus does not share the tensor structure of the result in two dimensions.

In fact prescription two turns out to be the \mbox{'t Hooft}-Veltman prescription introduced in Ref.~\cite{tHooft:1972tcz} and worked out in Ref.~\cite{Breitenlohner:1977hr}.
One can reproduce the result above by introducing the evanescent metric $\tilde{\eta}$:
\begin{align}
 \eta=\hat{\eta}+\tilde{\eta}\,.
\end{align}
Which means:
\begin{align}
 \tilde{\eta}\cdot \hat{\eta}&= 0\,,&\tilde{\eta} \cdot\tilde{\eta}&= \tilde{\eta}\,,&\tilde{\eta} \cdot\eta&= \tilde{\eta}\,.
\end{align}
With the corresponding index decomposition we find (employing the definition of $\gamma_3$):
\begin{align}
 \{\gamma_3,\gamma^{\hat{\mu}}\}&=0\,,&[\gamma_3,\gamma^{\tilde{\mu}}]&=0\,.
\end{align}
Applying these rules to  $T_{t_2}$ yields:
\begin{align}
\hat{T}_{t_2}&={\rm tr}\{\gamma_{3}\slashed{p}\gamma_{3}\slashed{q}\}={\rm tr}\{\gamma_{3}(\slashed{\tilde p}+\slashed{\hat p})\gamma_{3}(\slashed{\tilde q}+\slashed{\hat q})\}={\rm tr}\{(\slashed{\tilde p}-\slashed{\hat p})(\slashed{\tilde q}+\slashed{\hat q})\}\nonumber\\
&={\rm tr}\{\slashed{\tilde p}\slashed{\tilde q}\}-{\rm tr}\{\slashed{\hat p}\slashed{\hat q}\}=N_S\big[(\tilde{q}\cdot\tilde{p})-(\hat{q}\cdot\hat{p})\big]\nonumber\\
&=N_S\big[(q\cdot p) -2(\hat{q}\cdot\hat{p})\big]\,.
\end{align}
Comparing the two result $T_{t_2}$ and $\hat{T}_{t_2}$ makes clear that only $T_{t_2}$ obtained with the ``loose hat'' prescription obeys our initially stated requirement that one should obtain the continued trace result,
by calculating it for $\hat{N}_V$ dimensions and then continuing the result by replacing  $\hat{\eta}\rightarrow \eta$, without changing the structure of the result.
That means the fact that the two $\varepsilon_{N_V}$ tensors automatically eliminate evanescent structures is not just a feature, 
but is required for the cases, where the remaining external tensors of the trace cannot sustain an evanescent anti-symmetrizer on their own.

Unfortunately this is clearly not the case for the \mbox{'t Hooft}-Veltman prescription.
Here the algebra requires the presence of evanescent contributions.
This is because the algebra distinguishes between evanescent and intrinsically $\hat{N}_V$ dimensional index space at the level of reduced traces.
Such an additional burden can only be carried by an infinite dimensional Clifford algebra,
because the dimension of $\tilde{\eta}$ is given by $N_{\epsilon}=N_V-\hat{N}_V$ and is thus not an integer.

We conclude that we have to apply the ``loose hat'' prescription in order to reach our goal of evaluating the traces such that they retain the $\hat{N}_V$ dimensional structure.
As required benefit of the prescription we are allowed to (anti)-commute $\gamma_{N_V+1}$ and use $\gamma_{N_V+1}\cdot \gamma_{N_V+1} =1$ within each trace,
provided we do eliminate evanescent contributions in a consistent way, whenever their contribution spoils the (anti-)~commutativity of $\gamma_{N_V+1}$.
We are going to present an explicit example of the latter case in the next section.
 
\section{Restoring the (anti\mbox{-})~commutativity of $\gamma_{N_V+1}$}\label{SEC:SINGLEGAMMA5ANDEVANS}
In this section we will give an explicit example where evanescent contributions prevent a single $\gamma_{N_V+1}$ from (anti)-commuting freely in a fermion trace.
In order to keep the appearing expressions as short as possible we will work in $\hat{N}_V=2$ dimensions.
However, all obtained strategies and results can be generalized for any higher target dimension, as long as the maximal decomposition is known.

In the following we use:
\begin{align}
 \gamma_{3}=[\tilde{\varepsilon}_{2}]_{\hat{\mu}_1\hat{\mu}_2}\Gamma_{2}^{\hat{\mu}_1\hat{\mu}_2}\,.
\end{align}
We are going to investigate the result of the following trace:
\begin{align}
 T_{e_1}=&{\rm tr}\{\gamma_{3}\slashed p_1\slashed p_2\slashed p_3\slashed p_4\}\,.\nonumber
\end{align}
Plugging in the definition of $\gamma_{3}$ yields a trace that can be evaluated with the help of the decomposition for $T_6$ in Eq.~(\ref{EQ:T6Dec}).
Keeping the order of arguments in the trace like displayed in each term and not eliminating any evanescent structure yields for $T_{e_1}$ and all its cyclic re-orderings (we use {\tt FORM} to expand the terms):
\begin{align}
 T_{e_1}=&N_S\big[-(p_1\star p_2)(p_3\cdot p_4)
                  +(p_1\star p_3)(p_2\cdot p_4)
                  -(p_1\star p_4)(p_2\cdot p_3)\nonumber\\
                 &-(p_2\star p_3)(p_1\cdot p_4)
                  +(p_2\star p_4)(p_1\cdot p_3)
                  -(p_3\star p_4)(p_1\cdot p_2)
                   \big]\,.
\end{align}
We use the shorthand $p_i\star p_j=[\tilde{\varepsilon}_{2}]_{p_i p_j}$.
The result was expected to be invariant under cyclic re-orderings, because the generic $N_V$ dimensional trace is cyclic by definition.
However, calculating a naive anti-commuted version of $T_{e_1}$:
\begin{align}
 T_{e_2}=&-{\rm tr}\{\slashed p_1\gamma_{3} \slashed p_2\slashed p_3\slashed p_4\}\,,\nonumber
\end{align}
yields:
\begin{align}
 T_{e_2}=&N_S\big[-(p_1\star p_2)(p_3\cdot p_4)
                  +(p_1\star p_3)(p_2\cdot p_4)
                  -(p_1\star p_4)(p_2\cdot p_3)\nonumber\\
                 &+(p_2\star p_3)(p_1\cdot p_4)
                  -(p_2\star p_4)(p_1\cdot p_3)
                  +(p_3\star p_4)(p_1\cdot p_2)
                   \big]\,.
\end{align}
Because the sign of the last three terms of $T_{e_2}$ is opposite to the ones of $T_{e_1}$, we can conclude that $\gamma_3$ does not anti-commute when using the full $N_V$ dimensional result of the trace.
However, when we label the terms that originate from the expansion of the evanescent anti-symmetrizer $A_3$ with $\mathcal{O}_E$ we obtain:
\begin{align}
 T_{e_1}=&N_S\big[-(p_1\star p_2)(p_3\cdot p_4)
                  +(p_1\star p_3)(p_2\cdot p_4)
                  -(p_1\star p_4)(p_2\cdot p_3)\nonumber\\
                 &-(p_2\star p_3)(p_1\cdot p_4)\mathcal{O}_E
                  +(p_2\star p_4)(p_1\cdot p_3)\mathcal{O}_E
                  -(p_3\star p_4)(p_1\cdot p_2)\mathcal{O}_E
                   \big]\,,\\
 T_{e_2}=&N_S\big[-(p_1\star p_2)(p_3\cdot p_4)
                  +(p_1\star p_3)(p_2\cdot p_4)
                  -(p_1\star p_4)(p_2\cdot p_3)\nonumber\\
                 &+(p_2\star p_3)(p_1\cdot p_4)\mathcal{O}_E
                  -(p_2\star p_4)(p_1\cdot p_3)\mathcal{O}_E
                  +(p_3\star p_4)(p_1\cdot p_2)\mathcal{O}_E
                   \big]\,.
\end{align}
That means $\gamma_3$ anti-commutes through $\slashed p_1$ when we eliminate all evanescent contributions $\sim\mathcal{O}_E$.
One could also say $\gamma_3$ anti-commutes modulo $\mathcal{O}_E$ terms.
It turns out that for each step of anti-commutation one gets another set of terms that in the sum yield zero.
When we perform all possible anti-commutation steps we in total end up with three independent elimination equations:
\begin{align}
       [\tilde{\varepsilon}_{2}]_{\hat{\mu}_1,\hat{\mu}_2}A_3(p_1,\hat{\mu}_1,\hat{\mu}_2;p_2,p_3,p_4)=0\,,\\
       [\tilde{\varepsilon}_{2}]_{\hat{\mu}_1,\hat{\mu}_2}A_3(p_1,p_2,\hat{\mu}_1;p_3,p_4,\hat{\mu}_2)=0\,,\\
       [\tilde{\varepsilon}_{2}]_{\hat{\mu}_1,\hat{\mu}_2}A_3(p_1,p_2,p_3;p_4,\hat{\mu}_1,\hat{\mu}_2)=0\,.
\end{align}
These equations are known as Schouten-identities and the knowledge of the maximal decomposition allows us to generate all of them hiding inside the trace result.

We are free to choose terms that should be eliminated within each Schouten-identity.
A smart choice is to not eliminate the terms in $T_{e_1}$ that are left after the $\mathcal{O}_E\rightarrow0$ limit is performed.
For example we can choose the replacement:
\begin{align}
(p_2\star p_3)(p_1\cdot p_4)&\rightarrow (p_1\star p_3)(p_2\cdot p_4)-(p_1\star p_2)(p_3\cdot p_4)\,,\\
(p_2\star p_4)(p_1\cdot p_3)&\rightarrow (p_1\star p_4)(p_2\cdot p_3)-(p_1\star p_2)(p_3\cdot p_4)\,,\\
(p_3\star p_4)(p_1\cdot p_2)&\rightarrow (p_1\star p_4)(p_2\cdot p_3)-(p_1\star p_3)(p_2\cdot p_4)\,.
\end{align}
That means applying these replacement rules to $T_{e_1}$ (indicated by $|_E$), yields:
\begin{align}
 T_{e_1}\big|_{E}= T_{e_1}\big|_{\mathcal{O}_E\rightarrow 1}\big|_{E}= T_{e_1}\big|_{\mathcal{O}_E\rightarrow 0}\big|_{E}= T_{e_1}\big|_{\mathcal{O}_E\rightarrow 0}\,.
\end{align}
Applying them to the results of all possible cyclic permutation or/and anti-commutation versions of $T_{e_1}$, we find for all of them:
\begin{align}
 T_{e_i}\big|_{E}&= T_{e_1}\big|_{\mathcal{O}_E\rightarrow 0}\,.
\end{align}
We can also calculate all variants of
\begin{align}
 T_{e_{-1}}=&-{\rm tr}\{ \gamma_{3}\slashed p_4 \slashed p_3\slashed p_2\slashed p_1\}\,.\nonumber
\end{align}
With the above elimination equations we obtain
\begin{align}
 T_{e_{-i}}\big|_{E}=T_{e_1}\big|_{\mathcal{O}_E\rightarrow 0}\,.
\end{align}
This result is required,
because the trace result may not depend on the reading direction.
That means the result has to be invariant under Eq.~(\ref{EQ:NOReadingDirecion}).
At leading evanescents this is trivially granted, 
however at higher evanescent levels this restricts the solution of the decomposition to be one of the aforementioned mirror symmetric ones.
We have checked that the statements made here remain unchanged when working in higher dimensions $\hat{N}_V\in\{3,4\}$.
We also checked for $\hat{N}_V=2$ that similar eliminations work at the next to leading evanescent level that means including an evanescent anti-symmetrizer of rank $\hat{N}_V+2$.
Having no inversion symmetric decomposition for $T_{10}$ and $T_{12}$ at hand, we could only check the next to leading evanescent elimination for $\hat{N}_V=4$
without the inversion relation of $ T_{e_{i}} \leftrightarrow T_{e_{-i}}$.

We can conclude that a consistent elimination of evanescent structures in traces $T_N$ with a single $\gamma_{N_V+1}$ is necessary in order to retain the (anti\mbox{-})\-~commutativity of $\gamma_{N_V+1}$
 for $N\geq 2\hat{N}_V+2$  for even $\hat{N}_V$ and $N\geq \hat{N}_V+2$ for odd $\hat{N}_V$. 
Moreover, we justified the statement that one is allowed to (anti\mbox{-})\-~commute $\gamma_{N_V+1}$ and use the cyclicity of the trace beforehand as long as one sticks to a specific elimination convention.
In fact one does not need to generate the full set of elimination rules, one can just stick to a convenient normal form of a trace and drop its evanescent contributions afterwards.
Convenient means that the $|_{\mathcal{O}_E\rightarrow 0}$ operation eliminates sufficiently many terms, 
just to make sure that the surviving terms can be spared from all elimination replacements.
We can only claim that such a choice is always possible using a heuristic argument.
Explicit proofs of this statement would be very welcome.

As a final example we investigate the bilinear decomposition property of the product $\gamma_3 \gamma_{\mu}$ for $\hat{N}_V=2$
using our prescription of evaluating the trace.
We have:
\begin{align}
 \gamma_3 \gamma^{\mu}=[\tilde{\varepsilon}_2]_{\hat{\nu}_1 \hat{\nu}_2}\Gamma_2^{\hat{\nu}_1 \hat{\nu}_2}\Gamma_1^{\mu}\,.
\end{align}
In $N_V$ dimensions we have the complete decomposition ($\mathcal{O}_E\neq0$):
\begin{align}
 \Gamma_2^{\nu_1 \nu_2}\Gamma^{\mu}_1=2\Gamma^{\rho}_1A_2(\nu_1,\nu_2;\rho,\mu)+\Gamma_3^{\nu_1 \nu_2 \mu}\mathcal{O}_E\,.\label{EQ:G2G1Decompo}
\end{align}
The last term is an evanescent contribution.
We can now check the bilinear orthogonality using $T_6$ of Eq.~(\ref{EQ:T6Dec}):
\begin{align}
 T_{\Gamma_2\gamma \Gamma_2\gamma}=& {\rm tr}\left\{\gamma_3 \gamma_{\mu} \gamma_{\nu}\gamma_3\right\}=[\tilde{\varepsilon}_2]^{\hat{\rho}_1\hat{\rho}_2}[\tilde{\varepsilon}_2]^{\hat{\rho}_3\hat{\rho}_4}{\rm tr}\left\{\gamma_{\hat{\rho}_1} \gamma_{\hat{\rho}_2}\gamma_{\mu} \gamma_{\nu}\gamma_{\hat{\rho}_3} \gamma_{\hat{\rho}_4}\right\}=N_S \eta_{\mu \nu}\,.
\end{align}
Dropping the evanescent pieces  ($\mathcal{O}_E=0$) of the decomposition in Eq.~(\ref{EQ:G2G1Decompo}) after contracting with $[\tilde{\varepsilon}_2]$ has to yield the same result:
\begin{align}
  T_{\Gamma_1 \Gamma_1}=&[\tilde{\varepsilon}_2]_{\hat{\rho} \mu} {\rm tr}\left\{ \gamma_{\hat{\rho}} \gamma_{\hat{\sigma}}   \right\} [\tilde{\varepsilon}_2]_{ \nu \hat{\sigma}}=N_S \eta_{\mu \nu}\,.
\end{align}
We can also check the result dropping only a single $\mathcal{O}_E$ term using $T_4$ of Eq.~(\ref{EQ:T4Dec}):
\begin{align}
  T_{\Gamma_2\gamma \Gamma_1}=& {\rm tr} \left\{ \gamma_3 \gamma_{\mu}  \gamma_{\hat{\sigma}}  \right\} [\tilde{\varepsilon}_2]_{ \nu \hat{\sigma}}=N_S \eta_{\mu \nu}\,.
\end{align}
We see that in this case the evanescent contributions $\sim \mathcal{O}_E$ are automatically removed, 
because all three results are in agreement with the two dimensional one after the continuation $\hat{\eta}\rightarrow \eta$.

\section{Blueprint of a Reduction Prescription}\label{SEC:REDPRES}
We can now formulate how a fermion trace $T_N$ has to be evaluated in order to allow for a uniform dimensional continuation.
Note that the given prescription is just a blueprint.
That means it is not guaranteed to work for all possible cases.
\begin{enumerate}
 \item{If $\hat{N}_V$ is odd:
 Use Eq.~(\ref{EQ:OddToEvenTrace}) to obtain a trace that contains an even number of Dirac matrices.
 }
\item{If $\hat{N}_V$ is even: Discard all traces containing an odd number of Dirac matrices.}
 \item{ (Anti-)~Commute $\gamma_{N_V+1}$ to the left and use $\gamma_{N_V+1}\cdot \gamma_{N_V+1} =1$.\\ 
 After that one is left with traces that contain one or zero $\gamma_{N_V+1}$. }
 \item{ Use Eq.~(\ref{EQ:AntiCommutingThroughgammas}) to eliminate all trace internal contraction indices or multiple appearances of the same vector. 
 The elimination should be implemented in the same way like it is done in the \verb|trace4| function of {\tt FORM}.\footnote{The elimination is motivated by requirements imposed by renormalization and goes beyond the scope of this article.}
 }
 \item{  Bring the obtained trace in a normal form. 
 That means all variations of the trace that can be generated using the cyclicity, inverting the reading direction (see Eq.~(\ref{EQ:NOReadingDirecion})) and/or (anti\mbox{-})\-~commuting of the single $\gamma_{N_V+1}$ have to be mapped onto the same trace expression.}
  \item{ Use:
\begin{align}
 \gamma_{n+1}= i^{n(n-1)/2}[\varepsilon_{n}]_{\hat{\mu}_1\dots\hat{\mu}_{n}}\Gamma_{n}^{\hat{\mu}_1\dots\hat{\mu}_{n}}\,,
\end{align}
 to express the remaining $\gamma_{N_V+1}$ in terms of Dirac matrices.
  }
  \item{ Eliminate all $[\varepsilon_{n}]$ tensors originating from different traces and/or projectors using:
  \begin{align}
 [\varepsilon_{N_V}]_{\hat{\nu}_1,\dots,\hat{\nu}_{N_V}}~[\varepsilon_{N_V}]^{\hat{\mu}_1,\dots,\hat{\mu}_{N_V}}= \hat{N}_V! A_{N_V}(\hat{\nu}_1,\dots,\hat{\nu}_{N_V};\hat{\mu}_1,\dots,\hat{\mu}_{N_V})\,.
\end{align}
When more than three $[\varepsilon_{N_V}]$-tensors appear (including the ones in a potential projector), 
one has to perform the pairing in a standard way.
For example an average over all possible combinations of pairings does the job.
Left over $[\varepsilon_{N_V}]$ tensors have to be set to zero.
  }
 \item{Use the maximal decomposition of $T_N$ given in Eqs.~(\ref{EQ:T4Dec}-\ref{EQ:T8Dec}) to express the traces in terms of anti-symmetrizers. }
   \item{ Ensure that all problematic evanescent contributions originating from anti-symmetrizers $A_{N}$ with $N\geq \hat{N}_V+1$ do not contribute.}
  \item{ Expand the remaining $A_N$'s using their recursive definition of Eq.~(\ref{EQ:RECANDEF}) following the ``loose hat'' prescription of Eq.~(\ref{EQ:LooseHatPrescription1}-\ref{EQ:LooseHatPrescription3}).
  All internal contraction indices appearing in the maximal decomposition have to be treated to be $\hat{N}_V$ dimensional.
  All external indices and momenta which appear in propagators are to be treated continuous $N_V$-dimensional.
  }
 \end{enumerate}
 \newpage
 
\section{Checking the $VVA$-Anomaly}\label{SEC:VVAANOMALY}

\begin{figure}
\begin{center}
\begin{tabular}{cc}
  \includegraphics[scale=1.2]{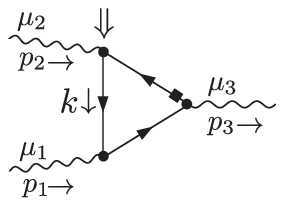}& 
  \includegraphics[scale=1.2]{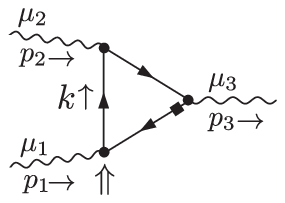}
\end{tabular}
\end{center}
\caption{\label{FG:FDVVA}Feynman diagrams relevant for the $VVA$ anomaly in $\hat{N}_V=4$ dimensions.
The propagator labeled with loop momentum $k$ does not carry any external momentum.
Black circles correspond to vector couplings $\sim \gamma^{\mu}$.
Black rectangles indicate the position of $\gamma_5$.
The double arrows indicate the starting point of the trace chosen in our setup.}
\end{figure}

We have implemented the reduction prescription in a self written {\tt FORM} program to perform checks.
One of the checks we went through is the calculation of the $VVA$ anomaly appearing for $\hat{N}_V=4$.
The anomaly is responsible for the non-conservation of the axial vector current $j_{AV}^{\mu}=\bar{\psi}\gamma_5\gamma^{\mu}\psi$ even in the case of massless fermions $\psi$, when one takes into account quantum corrections.
That means:
\begin{align}
 \partial_{\mu}j_{AV}^{\mu}\neq 0\,.\label{EQ:NONCSAVC}
\end{align}
independent of the mass of the fermions.
At classical level (no loop corrections) the axial vector current is conserved for vanishing fermion mass.
That means all terms appearing on the r.h.s.\ of Eq.~(\ref{EQ:NONCSAVC}) on classical level have an explicit fermion mass dependence.
For simplicity we discuss the anomaly only for the case of an abelian gauge theory, thus we are only dealing with the singlet axial vector current. 
In order to calculate the anomaly we evaluate the two Feynman diagrams in Fig.~\ref{FG:FDVVA} with the help of our {\tt FORM} program.
We keep all appearing couplings equal to one.
Because we only want to obtain the anomaly in momentum space we contract the amputated Green's function with $p_3^{\mu_3}=p_1^{\mu_3}+p_2^{\mu_3}$.
If the axial vector was conserved this should nullify the Green's function.
We keep the calculation as simple as possible and expand naively in small external momenta $p_1$ and $p_2$ and keep a large fermion mass $m$.
We can also assume the on-shell conditions $p_1^2=p_2^2=0$. Then all appearing integrals reduce to the class of massive one-loop tadpole integrals and can be evaluated in $N_V=d$ dimensions.
Employing a simple tensor reduction allows us to reduce scalar products like $k\cdot p_1$ and $k\cdot p_2$ to scalar products $p_1\cdot p_2$ times scalar products $k \cdot  k$.
The latter ones can be written in terms of inverse massive propagators.
To get rid of the tensor structure and to follow our prescription $\varepsilon_4\cdot \varepsilon_4\rightarrow A_4$ we use the projector:
\begin{align}
 \hat{P}^{\hat{\mu}_1\hat{\mu}_2}=\hat{\mathcal{N}}\tilde{\varepsilon}_4^{p_1p_2 \hat{\mu}_1\hat{\mu}_2}\,,
\end{align}
in order to directly project on the coefficient $\hat{c}_A$ of the anomaly structure
\begin{align}
 \hat{T}^{\hat{\mu}_1\hat{\mu}_2}=\tilde{\varepsilon}_4^{p_1p_2 \hat{\mu}_1\hat{\mu}_2}\,.
\end{align}
That means we keep the vector indices of external particles four dimensional.
The projector normalization is then given by $\hat{\mathcal{N}}=-2(p_1\cdot p_2)^{-2}$. 
Note that $\tilde{\varepsilon}_4$ is equivalent to {\tt e\_} used in {\tt FORM}.
When we follow the full trace calculation prescription and drop all mass dependent terms
after taking the small $2\epsilon=\hat{N}_V-N_V=4-d$ limit $\epsilon\rightarrow0$, we obtain:
\begin{align}
 \hat{c}_A=8 L=\frac{1}{2\pi^2}\,.
\end{align}
Where the loop factor is given by $L^{-1}=(4\pi)^{2}$.
One obtains the same result when using the {\tt FORM} function \verb|trace4| for the evaluation of traces.
This is a well known text book result (see for e.g. Eq.~({2.7.24}) of Ref.~\cite{Bohm:2001yx}).
In fact we can skip any of the steps three, four and five of the prescription and still obtain this result.
Skipping step four however, will require us to use a decomposition for $T_{10}$.
It is very important to notice that even in this case no evanescent terms $\sim \mathcal{O}_E$ stemming from an $A_5$ appear in $\hat{c}_A$.
We even can do any cyclic reordering of the trace before we plug in the decompositions.
This is because the two $\tilde{\varepsilon}_4$ tensors carrying $\hat{N}_V=4$ dimensional indices automatically eliminate any evanescent contributions in the trace.
Similar to the $\hat{T}_{t_2}$ example we already investigated.

We can go one step back and decide to use only continuous $N_V$ dimensional indices for the projector:
\begin{align}
 P^{\mu_1\mu_2}=\mathcal{N}\tilde{\varepsilon}_4^{p_1p_2 \mu_1\mu_2}\,.
\end{align}
This changes the projector normalization to:
\begin{align}
 \mathcal{N}=(1+\epsilon)(1+2\epsilon)\hat{\mathcal{N}}\,.
\end{align}
Still replacing $\gamma_5$ in the usual way and not changing its initial position within the trace (see Fig.~\ref{FG:FDVVA}) yields:
\begin{align}
c_A=0\,.
\end{align}
This is not the naively expected result, 
but it is the correct result one has to obtain when using a purely $N_V=d$ dimensional Clifford-algebra,
because we take into account all evanescent contributions through a ``bad'' positioning of $\gamma_5$ within the trace.
Labeling the terms stemming from $A_5$ of the $T_{10}$ decomposition with $\mathcal{O}_E$ indeed reveals:
\begin{align}
c_A=(1-1\cdot\mathcal{O}_E)L\,,
\end{align}
where both diagrams contribute in a fully symmetric way.
Thus the anomaly is only non-vanishing, when evanescent terms are properly dropped.
Therefore the discussed $VVA$ anomaly is an intrinsically four dimensional object.

However, sticking to the $d$-dimensional approach and anti-commuting $\gamma_5$
to the left of each trace (indicated by double arrows in Fig.~\ref{FG:FDVVA}) right in front of $\gamma_{\mu_2}$ in the first diagram and $\gamma_{\mu_1}$ in the second diagram
before using the trace decomposition one obtains again:
\begin{align}
 \hat{c}_A^{\prime}=8 L=\frac{1}{2\pi^2}\,.
\end{align}
The result is free of $\mathcal{O}_E$ terms and both diagrams still contribute in a fully symmetric way.
That means we eliminated all evanescent contribution just by performing a naive anti-commutation in each diagram.
In more detail we have chosen a position of $\gamma_5$ in the trace where the anti-symmetrizer $A_5$ cannot contribute, 
when its first quintet of indices is contracted through four indices with $\tilde{\epsilon}_4$ and one index with $\mu_1$ or $\mu_2$ (depending on the diagram).
But for its second quintet of vector indices we only have left $k,p_1,p_2 $ and $\mu_2$ or $\mu_1$.
That means we cannot saturate the second quintet of indices with five different sources and thus the $A_5$ contribution will vanish.
So through a smart positioning of $\gamma_5$ within a trace it is possible to automatically eliminate evanescent contributions.
This is in fact the approach Kreimer's prescription~\cite{Kreimer:1989ke,Korner:1991sx,Kreimer:1993bh} follows.
It is a very elegant approach, because it only requires the reduction of  Dirac chains in continuous $N_V=d$ dimensions.
However, to the best of the author's knowledge, it has not been proven that one can always eliminate any harmful evanescent contributions just by a naive anti-commutation of $\gamma_5$.
Especially for the case of purely virtual fermion traces -- where the lack of external particles makes a unique designation of a reading point quite difficult --
one can have doubts that there is always a configuration at hand which automatically cancels the evanescent contributions in the sum of all contributing diagrams.
Especially for the case  where ten different vectors or vector indices appear within a single trace.
Here there is clearly no positioning of $\gamma_5$ available such that $A_5$ does vanish within this trace
and the cancellation (if it does take place) has to happen in the sum of all traces.

We leave the check that the singlet $VVA$ anomaly does not receive radiative corrections beyond the one loop order for future work.
It would also be interesting to re-evaluate the results of Ref~\cite{Mondejar:2012sz} within our prescription.

Besides the $VVA$ anomaly analysis, we applied and checked our prescription in multi-loop $Z$-factor calculations, 
which we are going to discuss in the next section.

\section{Discussion}\label{SEC:DISCUSSION}
We are now in the position to discuss various prescriptions used in literature comparing to the prescription defined in this article.

Comparing the \verb|trace4| algorithm implemented in the computer algebra program {\tt FORM} to our prescription only yields different results for traces $T_N$ with  $N\geq 10$ and more than four different external indices/vectors.
Like the {\tt FORM} manual states it provides very short results. 
This means that it does not just discard the contribution of evanescent anti-symmetrizers appearing in the decomposition of the $N_V$ dimensional trace result, 
but in fact uses all available evanescent equations to eliminate as many terms as possible.
In fact \verb|trace4| does not do this explicitly but implicitly by the application of the four dimensional reduction of Eq.~(\ref{EQ:3GammaDecompoIn4D}).
However, the eliminations performed are not suitable for the renormalization program, because it eliminates too many terms of the traces.

Having identified evanescent contributions in a trace we are in the position to understand why the very ``ad hoc'' prescription introduced by Larin~\cite{Larin:1993tq}
does work.
In order to calculate the pure QCD corrections for the non-singlet piece of the axial-vector current:
\begin{align}
 J^{\mu}_{5 a}=\tfrac{1}{2}\bar\psi [\gamma_5,\gamma^{\mu}]t^a\psi\,.
\end{align}
He rewrites $\tfrac{1}{2}[\gamma_5,\gamma^{\mu}]$ in terms of $[\tilde{\varepsilon}_4]^{\mu \nu_1 \nu_2 \nu_3}\Gamma_{3,\nu_1 \nu_2 \nu_3}$
and uses a projector proportional to the same structure.
This reduces the calculation of the axial-vector current coefficient to a trace calculation.
Before evaluating the trace he uses Eq.~(\ref{EQ:AnSplitTo2eps}) to eliminate the two $\tilde{\varepsilon}_4$ tensors in favor of $A_4$
which is then treated to contain continuous $N_V$ dimensional $\eta$ tensors.
This then allows us to evaluate the trace in continuous $N_V$ dimensions.
Clearly, Larin's prescription does not induce an anti-commuting $\gamma_5$ and the fact that it does not anti-commute implies that it takes into account and includes $\hat{N}_V=4$ evanescent contributions,
because one evaluates the trace with an infinite dimensional Clifford algebra.
That means one has to evaluate traces $T_N$ with $N=10$ already at the one-loop level.
Because the evanescent contributions have a lower degree of divergence, they do not affect the renormalization of the leading UV poles.
Moreover, after performing the sub-divergence subtraction explicitly via insertion of explicit wave function and vertex counterterms,
they are only made finite due to the remaining overall divergence.
But since the overall divergence can be absorbed by a multiplicative renormalization constant $Z^{ns}_{MS}$,
all evanescent contributions can be subtracted by an inclusive finite renormalization constant $Z^{5}_{ns}$
which then restores the Ward-identity that connects the non-singlet vector current with the non-singlet axial vector current.
However, Larin's prescription of subtracting evanescent contributions at the integrated level with a finite correction factor cannot be straightforwardly applied to amplitudes which are not renormalized by a single, overall $Z$-factor,
once the sub-divergences have been subtracted. 
In this case the factorization ansatz for the finite renormalization cannot be used,
because different terms of the amplitude contain different amounts of evanescent contributions.
This is for example already the case when one considers massive fermions in the loops that require a mass renormalization.
Notice that the mass terms in the numerators of propagators undergo a completely different reduction and divergence power counting than the slashed momenta during the trace evaluations.
In any case removing evanescent contributions at the integrated level is to be considered a quite pragmatic approach and should be abandoned as soon
as one gets a handle on them at the integrand level.

Our prescription justifies the approach to anti-commute $\gamma_5$ and obtained $\varepsilon_4$-tensors like in four dimensions 
for the evaluation of $\beta$-functions in the SM up to including three loops in Refs.~\cite{Mihaila:2012fm,Mihaila:2012pz,Chetyrkin:2012rz},
because the divergent pieces of the three-loop calculation are not sensitive to evanescent contributions when following either our prescription or simply using the function $\verb|trace4|$ in {\tt FORM}
to reduce the appearing traces.
We checked the absence of evanescent contributions
at three-loop level in the $Z$-factors required for the calculation of $\beta_{\alpha_s}$ the strong coupling $\beta$-function,
the $\beta$-function of the Yukawa-coupling for top $\beta_{\alpha_t}$ and bottom quark $\beta_{\alpha_b}$,
as well as the Higgs self-coupling $\beta_{\lambda}$ by an explicit calculation, taking into account all corrections proportional to $\alpha_s,\alpha_t,\alpha_b$ and $\lambda$.

We can also comment on the recent evaluation of the four-loop Yukawa $\beta$-function in Refs.~\cite{Bednyakov:2015ooa,Zoller:2015tha}.
Both groups use Kreimer's prescription~\cite{Kreimer:1989ke,Korner:1991sx,Kreimer:1993bh} for dealing with $\gamma_5$ within DREG.
The first group also uses the \mbox{'t Hooft}-Veltman prescription elaborated in Ref~\cite{Breitenlohner:1977hr} in order to validate the problematic non-naive contribution.

The obtained results by both groups indicate reading point dependent results for the $\beta$-function.
That means different places where $\gamma_5$ is naively anti-commuted to before the trace evaluation takes place yield different results.
There already appeared publications~\cite{Poole:2019txl,Poole:2019kcm} 
about how to resolve this intrinsic ambiguity with the help of Weyl consistency conditions obtained from curved space-time considerations.\\
With the insight obtained in this article, we suggest, that the reading point dependence could arise due to the missing elimination of evanescent contributions,
which we checked already appear at the three-loop level in finite parts of the amplitudes involved.
There we start to run into traces $T_N$ with $N\geq 10$ with six different trace external momenta/indices and a single $\gamma_5$ when following our reduction prescription.
However, for traces $T_N$ with $N< 10$ our prescription yields the same results as Kreimer's prescription when calculating $\beta$-functions.
The major difference is that our prescription does not attempt to embed the four dimensional algebra in a continuous $N_V$ dimensional one,
but continues the trace external momenta and indices -- appearing in the trace result -- to continuous $N_V$ dimensions, 
trying to avoid the contamination of $\hat{N}_V$ evanescent contributions.

Last but not least we understand why \verb|trace4| is applicable for a $Z$-factor calculation within the chiral-XY Gross-Neveu-Yukawa model up to including four loops.
Simply because there are not sufficiently many ($N<10$) Dirac matrices in the trace $T_N$ to create any evanescent contributions for $\hat{N}_V=4$.

\section{Summary}
In this article we have been investigating how the results of fermion trace evaluations in integer dimensions $\hat{N}_V$ can be continued to non-integer dimensions $N_V$
in a uniform way. Such a continuation is required for a dimensional regulator in the presence of $\gamma_{N_V+1}$.

This approach is conceptually very different to the one where one lifts the $so_N$ and Clifford algebra from integer dimensions to generic non-integer dimensions like this is done in DREG.
We claim that the latter only works when intrinsically integer dimensions contributions like the ones $\sim\varepsilon$ are allowed to be dropped.
In order to distinguish both methods we call regulators employing dimensional continuation Dimensionally Continued Regularization (DCREG).

We identified evanescent contributions appearing in fermion traces $T_N$ for $N\geq \hat{N}_V$
and noticed that their presence can stop the intrinsic integer dimension Dirac matrices $\gamma_{N_V+1}$ from (anti)-commuting.
In order to allow a uniform continuation and respect the (anti\mbox{-})\-~commutativity of $\gamma_{N_V+1}$ a systematic elimination of these evanescent contributions has to be carried out.
We worked out that the proper treatment of $\hat{N}_V$ and $N_V$ dimensional vector indices is given by the loose hat prescription,
which does not correspond to the \mbox{'t Hooft}-Veltman prescription elaborated in Ref.~\cite{Breitenlohner:1977hr}.

A successful construction of a dimensional regulator that does not need symmetry restoring finite renormalization factors
requires full control over all present evanescent contributions.
We do not claim that the decomposition of fermion traces is sufficient to achieve this goal for any amplitude,
but showed that it is a relevant ingredient.

The performed application of our prescription up to the three loop level in the SM and four loop level in the chiral-XY Gross-Neveu-Yukawa model
could indicate that a symmetry preserving, dimensional regulator exists and an all order formulation might be in reach.
The steps performed in this paper can thus be understood as very basic steps on the way to a dimensional solution of the ``$\gamma_5$ problem''.

However, we leave the explicit elimination of evanescent structures which are beyond the loose hat prescription 
and all its impact on an actual multi-loop calculation, 
like in the determination of the SM $\beta$-functions at the four-loop level and the calculation of radiative corrections to the $VVA$ anomaly for future work.
Of course this also includes the explicit discussion of the constraints on the evaluation of traces due to renormalization.
One further major obstacle is the determination of complete and symmetric decomposition of $T_N$ with higher values of $10 \leq N < 20$.

\section{Acknowledgment}
We thank Oliver B\"ar, Joachim Brod, Martin Gorbahn, John Gracey, Manfred Kraus, Dirk Kreimer, Luca Di Luzio,  Peter Marquard, Ayan Paul and Bas Tausk for useful discussions.
We thank Andreas Maier for valuable contributions concerning an efficient reduction of multi-loop amplitudes in FORM.
We thank Sascha Peitzsch and Peter Marquard for reading the full manuscript.

\end{document}